\documentclass[journal,draftcls,onecolumn,10pt,twoside]{IEEEtran}

\usepackage{hyperref}

\usepackage{texPreamble}
\usepackage{texJamesPreamble}
\usepackage[numbers,sort&compress]{natbib}

\usepackage{geometry}
\IEEEoverridecommandlockouts

\DeclareMathOperator{\Vol}{Vol}

\begin{document}

\title{New Bounds on the Size of Binary Codes with Large Minimum Distance}

\author{
James (Chin-Jen) Pang, Hessam Mahdavifar, and S. Sandeep Pradhan  

\thanks{
This work was supported by the National Science Foundation
under grants CCF--2132815 and CCF--1909771.
This paper was presented in part at the 2022 IEEE International Symposium on Information Theory.
}%
\thanks{The authors are with the Department of Electrical Engineering and Computer Science, University of Michigan, Ann Arbor, MI 48109, USA (e-mail: cjpang@umich.edu; hessam@umich.edu; pradhanv@umich.edu).}
}

\maketitle

\begin{abstract}
Let $A(n, d)$ denote the maximum size of a binary code of length $n$ and minimum Hamming distance $d$. Studying $A(n, d)$, including efforts to determine it as well to derive bounds on $A(n, d)$ for large $n$'s, is one of the most fundamental subjects in coding theory. In this paper, we explore new lower and upper bounds on $A(n, d)$ in the large-minimum distance regime, in particular, when $d = n/2 - \Omega(\sqrt{n})$. 
We first provide a new construction of cyclic codes, by carefully selecting specific roots in the binary extension field for the check polynomial, with length $n= 2^m -1$, distance $d \geq n/2 - 2^{c-1}\sqrt{n}$, and size $n^{c+1/2}$, for any $m\geq 4$ and any integer $c$ with $0 \leq c \leq m/2 - 1$. These code parameters are slightly worse than those of the Delsarte--Goethals (DG) codes that provide the previously known best lower bound in the large-minimum distance regime. However, using a similar and extended code construction technique we show a sequence of cyclic codes that improve upon DG codes and provide the best lower bound in a narrower range of the minimum distance $d$, in particular, when $d = n/2 - \Omega(n^{2/3})$.
Furthermore, by leveraging a Fourier-analytic view of Delsarte’s linear program, upper bounds on $A(n, \ceil{n/2 - \rho\sqrt{n}\,})$ with $\rho\in (0.5, 9.5)$ are obtained that scale polynomially in $n$. To the best of authors' knowledge, the upper bound due to Barg and Nogin \cite{barg2006spectral} is the only previously known upper bound that scale polynomially in $n$ in this regime. We numerically demonstrate that our upper bound improves upon the Barg-Nogin upper bound in the specified high-minimum distance regime.
\end{abstract}

\section{Introduction}\label{sec:Intro}
An error-correcting code $C$ of length $n$ and minimum distance $d$ over a finite field $\Fq$ is a subset of the vector space $\Fq^n$ with $d = \min d_H(x, y)$, over all distinct $x, y\in C$. Here, $d_H(x, y) = \sum_{i=1}^n \mathbbm{1}_{\mathset{x_i \neq  y_i}}$ is the Hamming distance between $x$ and $y$. The code $C$ is said to be linear if $C$ is a subspace of the vector space $\Fq^n$. 
The capabilities and limitations of error-correcting codes are, in general, closely related to their minimum distance. For instance, the maximum number of errors a code can correct in the Hamming space is upper bounded by half of its minimum distance. This has led to a vast range of studies spanning several decades to answer one of the most fundamental and classical problems in coding theory, which is to determine (or to derive bounds on) the maximum size $A_q(n,d)$ of an error-correcting code $C$ of length $n$ over $\Fq$ and with minimum distance $d$ \cite{macwilliams1977theory, guruswami2012essential, van1998introduction}. 
Several of the most well-known results in the literature focus on the regime where $n\goto \infty$ and $d$ is proportional to $n$, namely $d = \delta n$, for some $0< \delta < 1$.
The question then is to find the asymptotic maximal rate $R(\delta)$ of an error-correcting code with relative distance $\delta$, where we define $R(\delta) \deff \limsup_{n\rightarrow \infty} \frac{1}{n} \log_q A_q(n,  \floor{n\delta})$.

Lower bounds on $A_q(n, d)$ are often obtained by constructions, either explicitly or implicitly, i.e., via existence arguments. One of the most well-known lower bounds on $A_q(n,d)$ 
is the Gilbert--Varshamov (GV) bound \cite{gilbert1952comparison,varshamov1957estimate}:
\[
A_q(n,d) \geq \frac{q^n}{V_q(n ,d-1)}, 
\] where $V_q(n ,d) = \sum_{i=0}^d \binom{n}{i}(q-1)^i$ is size of a Hamming ball of radius $d$ in $\Fq^n$.
Several improvements have been proposed to strengthen the GV bound, see, e.g., \cite{elia1983some, barg2000strengthening, o2006bounds, jiang2004asymptotic, spasov2009some, ye2021improving}. Among them, the most notable improvement in the binary case is due to Jiang and Vardy \cite{jiang2004asymptotic}, who improved the GV lower bound on $A_2(n,d)$ for $\delta < 0.499$ by a multiplicative factor of $c\log_2 V_2(n ,d)$, for some constant $c$, via studying the independence number of the sparse Gilbert graph on $\Ftwo^n$. 
For prime powers $q = p^{2k}$ with $q \geq 49$, explicit constructions of $q$-ary linear codes, obtained through algebraic-geometric codes, that surpass the GV bound are known \cite{tsfasman2013algebraic}. 
For $q=2$, a well-known conjecture asserts that the binary version of the GV bound is asymptotically tight, when expressed as a lower bound on $R(\delta)$. For a survey on the known bounds with finite $n$ and $d$, the reader is referred to \cite{litsyn1998update} and the websites \cite{agrell, brouwer}. 
For asymptotic lower bounds and an overview of known results the reader is referred to \cite{jiang2004asymptotic, gaborit2008asymptotic}.

The best asymptotic upper bounds currently known are due to McEliece, Rodemich, Ramsey and Welch (MRRW) \cite{mceliece1977new}. Built upon Delsarte’s linear program (LP) approach \cite{delsarte1973algebraic}, these bounds are established by showing valid solutions to the dual LPs, and are often called the first and the second linear programming bounds. 
In addition to the original proof in \cite{macwilliams1977theory}, which utilizes Delsarte's LP and properties of Krawtchouk polynomials, the first LP bound has also been proved using various techniques including harmonic analysis of Boolean functions \cite{friedman2005generalized, navon2009linear, samorodnitsky2021one}, spectral analysis \cite{barg2006spectral}, and functional and linear-algebraic approaches  \cite{barg2008functional}.
{There is substantial empirical evidence \cite{barg1999numerical} indicating that the bounds in \cite{mceliece1977new} asymptotically give the exact answer in the asymptotic Delsarte problem}.
Consequently, several works introduce new hierarchies of LPs, which include Delsarte’s LP as the {weakest member of this family} \cite{coregliano2021complete, loyfer2022linear, loyfer2022new}. Among them, the concurrent works \cite{coregliano2021complete} and \cite{loyfer2022linear} study similar families of LPs applicable to linear codes only, and empirically show significant improvement compared to Delsarte's. In \cite{loyfer2022new}, new hierarchies of LPs for linear and general codes along with the first dual feasible solutions to the LP’s that recover the first LP bound, are developed.

\subsection{Motivation}\label{subsec:Motivation}
Low-rate codes are becoming increasingly important with the emergence of low-capacity scenarios, including the Internet of Things (IoT) and satellite communications. For instance, in IoT network, the devices need to operate under extreme power constraints and often need to communicate at very low signal-to-noise ratio  \cite{ratasuk2016overview}.
In the standard, legacy Turbo codes or convolutional codes at moderate rates together with many repetitions are adopted to support communication at low rates. It is expected, however, that repeating a moderate-rate code to enable low-rate communication will result in rate loss and suboptimal performance. 
As a result, studying low-rate error-correcting codes for reliable communications in such low-capacity regimes has become a subject of extensive recent works \cite{shirvanimoghaddam2016raptor,fereydounian2019channel,abbasi2021hybrid, dumer2020codes, abbasiy2021polar, dumer2021combined, abbasi2022polar, jamali2021massive, dumer2021codes, vaezi2022cellular, abbasi2022hybrid}. 

Motivated by the need to revisit various aspects of channel coding in the low-rate regime, from efficient code design to reliable decoding algorithms, in this paper we focus on the minimum distance properties of binary codes in the low-rate regime, which can be also described as the \text{large} minimum distance regime, to be specified later. Let $C$ be a binary code of length $n$, size $M$, and minimum distance $d = (n - j)/2$, referred to as an $(n, M, d)$ code (for the sake of simplifying the equations, we reserve the parameter $j$ to denote $n-2d$ in various places throughout the paper). With a slight abuse of terminology, the \textit{dimension} of $C$, including for non-linear codes, is denoted by $k = \log_2 M$. Also, let $R = k/n$ denote the code rate. 
In this paper, we focus on studying bounds on $A_2(n, d)$ in the large-minimum distance regime, in particular, when $d = n/2 - \Omega(\sqrt{n})$, i.e., $j = \Omega(\sqrt{n})$. 
For ease of notation, we use $A(n, d)$ to denote $A_2(n,d)$ throughout the paper keeping in mind that the focus is on studying binary codes.

\subsection{Related Works}\label{subsec:intro_related}
For $j=n-2d \leq 0$, provided that a sufficient number of Hadamard matrices exist, a widely accepted conjecture, Plotkin and Levenshtein (see \cite[Chapter 2, Theorem 8]{macwilliams1977theory}) have essentially settled the problem and showed that $A(2d, d) =4d $, $A(n,d) =2\floor{d/ (2d-n)}$ for even $d > n/2$, and $A(n,d) =2\floor{\frac{d+1}{2d+1- n} }$ for odd $d > (n-1)/2$.

In what follows, we consider the scenario with $j >0$. When $j$ scales linearly with $n$, asymptotic results can be found in \cite{guruswami2012essential, jiang2004asymptotic}. In particular, the conjecture is that there does not exist any binary code exceeding the GV lower bound (Theorem \ref{Thm:GV}).
There are a limited number of studies in the literature targeting the regime where $j$ is sub-linear in $n$. In 1973, McEliece (see \cite[Chapter 17, Theorem 38]{macwilliams1977theory}), utilizing the LP approach, established the following bound that is valid for $j = o(\sqrt{n})$:
\begin{equation}\label{eq:ME_bound}
    A(n, d) \lesssim n(j+2).
\end{equation}
For $j \approx n^{1/3}$, codes have been constructed \cite{sidel1971mutual} to meet McEliece's upper bound, and hence, showing the tightness of this bound in this regime.
A few improvements \cite{tietavainen1980bounds, krasikov1997upper} have been derived in the literature in the regime $j = o(n^{1/3})$. 
For $j = \Omega(\sqrt{n})$, the Delsarte–Goethals (DG) codes, first introduced as a generalization of Reed-Muller codes \cite{kasami1968new, delsarte1970generalized, goethals1974two, goethals1976nonlinear, hergert1990delsarte}, are a class of nonlinear code and are known to be the best known codes, in terms of the minimum distance given a code size, in this regime. When $j = \Theta(\sqrt{n})$, a sequence of DG codes with sizes scaling polynomially in $n$ can be constructed.

While no explicit upper bounds on $A(n,d)$ are derived in the literature targeting the specific regime $j = \Omega(\sqrt{n})$, the results by Barg and Nogin \cite{barg2006spectral} can be tailored to provide a sequence of bounds on $A(n,d)$ scaling polynomially in $n$ for $j = \Theta(\sqrt{n})$. To the best of our knowledge, this is the only existing result in the literature leading to upper bounds that scale polynomially with $n$ in this regime.
In this paper, we attempt to answer the following question: 
If the term $j = n-2d$ scales as $j = \Omega(\sqrt{n})$, what is the best size $M$ one can achieve?

\subsection{Our Contribution}\label{subsec:intro_contribution}
We study the cardinality  $A(n, d)$ of binary codes in the large minimum distance regime where $j = n-2d$ scales as $j = \Omega(\sqrt{n}) \cap o(n)$. In particular, we show the following results: 
\begin{itemize}
 \item Two code constructions with sizes scaling polynomially and quasi-polynomially in $n$ are presented for cases with $d = n/2 -\Omega(\sqrt{n})$ and $d = n/2 -\Omega(n^{2/3})$, respectively, by demonstrating explicit and carefully designed BCH-like cyclic linear codes.
    Specifically, for $c\in \N$, the first construction has size $n^{c+\frac{1}{2}}$ and $d \geq n/2 - 2^{c-1}\sqrt{n}$, and the second construction has size $n^{\frac{\log n}{6} + \frac{3c^2}{2} +\frac{5c}{2} +\frac56}$ and $d \geq n/2 - 2^{c-1} n^{2/3} - 2^{2c-1} n^{1/3}$. 
    \item 
    
    Compared with the state-of-the-art lower bounds on $A(n,d)$ based on the Delsarte--Goethals codes, the first cyclic construction is inferior by a multiplicative factor of $\Theta(n^{3/2})$ in the regime $j = \Theta(\sqrt{n})$.  
    In the regime $j = \Theta({n}^{2/3})$,  the second construction is superior to the DG codes by a multiplicative factor of $\Theta(n^{\frac32 c^2 + \frac32 c -1})$ and provides the best lower bound in this regime.
    \item Asymptotic upper bounds for $A(n, n/2 - \rho \sqrt{n})$, based on an improved bounding technique inspired by \cite{navon2009linear} and a new method to bound the maximal eigenvalues of adjacency matrix induced by a Hamming ball $B_r \in \bit^n$ with finite $r$, are shown. 
    \item The asymptotic scaling behaviour of the proposed Fourier-analytical based upper bounds for $A(n, n/2 - \rho \sqrt{n})$ and the spectral-based bounds derived from \cite{barg2006spectral}, both of which are polynomial in $n$, are plotted for $\rho \in (0.5, 9.5)$, where the former are slightly stronger. 
\end{itemize}

\subsection{Outline of the Paper}\label{subsec:outline}
The rest of this paper is organized as follows. In Section \ref{sec:Prelim} we review several well-known bounds on $A(n,d)$ and examine their scaling behaviour when $j = \Theta(\sqrt{n})$. Results in a prior literature by Barg and Nogin \cite{barg2006spectral} that can be used to provide upper bounds on $A(n, d)$ when $j = \Theta(\sqrt{n})$ are discussed in Section~\ref{subsec:prelim_Sasha}.
In Section \ref{subsec:Main_CyclicConstruciton}, a BCH-like cyclic code construction, with $j$ scaling from $\Theta(\sqrt{n})$ to $\Theta(n)$, is presented.
Section \ref{subsec: cyclic_general} describes another construction with better performance in the regime $j = \Omega(n^{\frac23})$.
In Section \ref{subsec:Upper_Navon}, we review an alternative proof of the {first linear programming bound} on $A(n,d)$ (formally decribed in Section~\ref{subsec:PrelimMRRW}) through a covering argument using Fourier analysis on the group $\Ftwo^n$. 
An asymptotic upper bound on $A(n,d)$ with $ d\geq n/2 - \sqrt{n}$ that are strictly tighter than all prior results is derived in Section~\ref{subsec:Upper_New_r3}. The upper bounding technique is extended in Section~\ref{subsec:Upper_general_r} and yields a family of bounds on $A(n,d)$ with $ d\geq n/2 - \rho \sqrt{n}$ for $\rho \in (0.5, 9.5).$ The bounds are compared with a sequence of bounds derived from \cite{barg2006spectral} for the same range of $d$ in Section~\ref{subsec:Upper_general_r_numerical}.
Finally, the paper is concluded in Section\,\ref{sec:Conclusion}. 

\section{Preliminaries}\label{sec:Prelim}
\textbf{Notation.} 
Let $f$ and $g$ be two real-valued functions of $n\in \N$. We write $f(n) \lesssim g(n)$ if $f(n) \leq \left(1+ o(1)\right)g(n)$, write $f(n) \gtrsim g(n)$ if $f(n) \geq \left(1+ o(1)\right)g(n)$, and write $f(n) \sim g(n)$ if $\lim_{n\goto \infty} f(n)/g(n) =1$.
Let $H_2(\cdot)$ denote the binary entropy function.
For positive integers $r, n \in \N$ with $n \geq r$, let $B_r(\zero,n) \in \bit^{n}$ denote the Hamming ball of radius $r$ centered at $\zero = (0, 0, \ldots, 0)$, and its volume by $\Vol(r, n) \deff$  $\abs{B_r(\zero,n) } = \sum_{i=0}^r \binom{n}{i}$.
When $n$ is clear from the context, we write $B_r$ and  $B_r(\zero,n)$ interchangeably. 
We recall the following bounds for $r\leq n/2$
\begin{enumerate}
    \item $\Vol(r,n) \leq 2^{H_2(r/n) n}$; and 
    \item $\Vol(r,n) \geq 2^{H_2(r/n) n - o(n)}$ for sufficiently large $n$.
\end{enumerate}

We study asymptotic lower and upper bounds on $A(n,d)$ in this section, and evaluate them in the large minimum distance regime $j = \Omega(\sqrt{n}) \cap o(n)$. 
\subsection{Lower Bounds}\label{subsec:Prelim_known}

We review some asymptotic lower bounds on $A(n,d)$ in this section. The first one is the well-known GV lower bound. Note that there is an improvement to the GV bound by Jiang and Vardy \cite{jiang2004asymptotic} that is not considered here because the constraint on the relative distance $0\leq \delta < 0.499$ in \cite{jiang2004asymptotic} does not hold for large $n$ when $j = \Omega(\sqrt{n}) \cap o(n)$. 
\begin{theorem}[GV lower bound, \cite{gilbert1952comparison, varshamov1957estimate}]\label{Thm:GV}
    Let positive integers $n$ and $d \leq n/2$ be given. Then 
    \begin{equation}\label{Eq:GV_nonAsym}
        A(n,d) \geq \frac{2^n}{\Vol(d-1, n) }.
    \end{equation}
\end{theorem}

Asymptotically, suppose $0\leq \delta < 1/2$, then there exists an infinite sequence of $(n, M, d)$ binary linear codes with $d/ n > \delta$ and rate $R = k/n$ satisfying 
        $R \geq 1 - H_2(d/n).$
To evaluate \Tref{Thm:GV} when $j = \Theta(\sqrt{n})$, consider $j= 2a \sqrt{n}$.
The central limit theorem, coupled with the Berry–Esseen theorem, provides an upper bound 
$\Vol(d-1, n)
= 2^n \left[ Q(2a) + O(1/\sqrt{n})\right],$
where $Q(\cdot)$ denotes the tail distribution function of the standard normal distribution. 
Hence we have
\begin{equation}\label{eq:GV_sp}
A(n, n/2-a\sqrt{n}) \geq \left[ Q(2a) + O(1/\sqrt{n})\right]^{-1},    
\end{equation}
which is loose compared with the Plokin-Levenshtein bound $A(2d, d) = 4d$.

The Delsarte–Goethals (DG) codes are a class of nonlinear codes that are associated with the Reed-Muller codes and are the best known codes for their parameters. 
\begin{theorem}[Delsarte--Goethals code
 \cite{kasami1968new, delsarte1970generalized, goethals1974two, goethals1976nonlinear, hergert1990delsarte}]
Let $m \geq 4$ be an even integer and $0 \leq r \leq {m}/{2} -1$ be an integer. The Delsarte--Goethals code DG$(m,r)$ is a binary code of block length $n=2^m$, size $2^k$, where $k = r(m-1) + 2m$, and minimum distance $2^{m-1} - 2^{m/2 +r -1}$. 
For $r  = {m}/{2} -1$, DG$(m,r) =$ RM$(m,2)$, the second order Reed-Muller code. For $0\leq r \leq {m}/{2} -2$,  DG$(m,r)$ is a nonlinear subcode of RM$(m,2)$. 
\end{theorem}
For the case when $j = \Theta(\sqrt{n})$, one may consider DG$(m,r)$ codes with a finite $r$, and show that $A(n,d) \geq 2^{-r} n^{r+2}$ for $n$ an even power of $2$ and $d ={n}/{2} - 2^{r-1}\sqrt{n}$. 
For the case when $j = \Theta(n^{2/3})$, considering DG$(m,r)$ codes with $m$ a multiple of $6$ and $r = m/6 +c$ for some finite $c$, one may show that $A(n,d) \geq n^2 \parenv{n/2}^{\frac{\log n}{6} +c}$, where $n= 2^{6\ell}$ for some $\ell \in \N$ and $d ={n}/{2} - 2^{c-1}n^{2/3}$.

\subsection{Asymptotic Upper Bounds}\label{subsec:PrelimMRRW}

The following upper bounds on the size of binary codes can be found in standard coding theory textbooks, e.g. \cite{macwilliams1977theory},\cite{guruswami2012essential}. Bounds for the regime $j = n-2d= \Theta(\sqrt{n})$ are derived and given following the general bounds, e.g. inequalities \eqref{eq:Hamming_sp},  \eqref{eq:Plotkin_sp}, \eqref{eq:Plotkin_sp2},  and \eqref{eq:EB_sp}. 
When the scaling behaviour of $j$ matters, we choose $j = 2a \sqrt{n}$, i.e., $d = n/2 -a\sqrt{n}$, for ease of comparison between bounds.

\begin{theorem}[Hamming Bound]\label{Thm:Hamming}
For every $(n, M, d)$ code  $C \subset \bit^{n}$, 
\begin{equation}\label{Eq:Hamming}
    M \leq 2^n/ \Vol(e, n),
\end{equation}
where $e = \floor{(d-1)/2}$.
\end{theorem}

In the asymptotics, \Tref{Thm:Hamming} bounds the rate from above, in terms of the relative distance $\delta$, by
$R \lesssim 1- H_2(\delta / 2).$
For $j = \Theta(\sqrt{n})$, the term $e = n/4 - \Theta(\sqrt{n})$, and 
$ \Vol(e, n) \geq 2^{H_2(1/4) n - o(n)}$. 
Hence Theorem \ref{Thm:Hamming} becomes 
\begin{equation}\label{eq:Hamming_sp}
    M \leq 2^{ (1 - H_2(1/4) )n + o(n) } \lesssim 2^{0.189 n}, 
\end{equation} for all sufficiently large $n$.

\begin{theorem}[Singleton Bound]\label{Thm:Singleton}
Let $C \subset \bit^{n}$ be a binary code with distance $d$ and dimension $k$, then $k \leq n-d +1.$
\end{theorem}
For $j = \Theta(\sqrt{n})$, \Tref{Thm:Singleton} yields $    M \leq 2^{n/2 +O(\sqrt{n})}, $
which is weak compared to \eqref{eq:Hamming_sp}.

\begin{theorem}[Plotkin Bound, \cite{plotkin1960binary}]\label{Thm:Plotkin}
The following holds for any code $C \subset \bit^{n}$ with distance $d$.
\begin{enumerate}
    \item If $d = n/ 2$, $\abs{C} \leq 2n$. 
    \item If $d > n/2$,  $\abs{C} < 2 \ceil{ \frac{d}{2d - n}}$. 
\end{enumerate}
\end{theorem}
One may use a combinatorial argument and \Tref{Thm:Plotkin} to derive the following corollary.
\begin{corollary}\label{Coro:Plotkin}
If a $(n, M, d)$ binary code $C$ has distance $d < n/2$,
then the size $M \leq d\cdot 2^{n- 2d +2}$.
\end{corollary}
Using \Cref{Coro:Plotkin}, one may bound the size of any code with $d = (n -j)/2 <n/2$ by \begin{equation}\label{eq:Plotkin_sp}
    M \leq d \cdot 2^{j +2} < 2n \cdot 2^j.
\end{equation}
When $j$ scales as $j = \Theta(\sqrt{n})$, the size $M$ is bounded sub-exponentially in $n$. In particular, set $j = 2a\sqrt{n}$, i.e. $d = \ceil{n/2 - a\sqrt{n}\,}$, \eqref{eq:Plotkin_sp} becomes 
\begin{equation}\label{eq:Plotkin_sp2}
    M \leq 2n \cdot 2^{2a\sqrt{n}}.
\end{equation}

\begin{theorem}[Elias-Bassalygo Bound]\label{Thm:EB}
For sufficiently large $n$, every code $C \subset \bit^{n}$ with relative distance $\delta\leq 1/2$  and rate $R$ satisfies the following:
$    R \lesssim 1- H_2(J_2(\delta)),$
where  $J_2(\delta) \deff \frac{1}{2}(1- \sqrt{1- 2\delta} )$.
\end{theorem}
Assuming $d = \ceil{ n/2 - a \sqrt{n}\, }$, one may adopt steps similar to the proof of Theorem \ref{Thm:EB} as in \cite[p.147]{guruswami2012essential} to show an upper bound:
\begin{equation}\label{eq:EB_sp}
    M \leq n^3 \cdot 2^{\frac{a}{\ln{2}}\sqrt{n} + O(1)}.
\end{equation}

The last upper bound we introduce is known as the \textit{first linear programming bound} or the \textit{MRRW bound} on binary error correcting codes, or, alternatively, on optimal packing of Hamming balls in a Hamming cube.
The bound was originally proved by McEliece, Rodemich, Rumsey, and Welch \cite{mceliece1977new}, following Delsarte’s linear programming approach \cite{delsarte1973algebraic}, and  is the best known asymptotic upper bound on the cardinality of a code with a given minimal distance scaling linearly in $n$, for a significant range of the relative distance.

\begin{theorem}[MRRW Bound, \cite{mceliece1977new}]\label{Thm:MRRW}
For sufficiently large $n$, every code $C \subset \bit^{n}$ with relative distance $\delta$  and rate $R$ satisfies the following:
\begin{equation}\label{eq:MRRW_2}
R \lesssim  H_2\left(1/2 - \sqrt{\delta(1- \delta)}\right). 
\end{equation}
\end{theorem}

\begin{remark}\label{rem:secondLPbound}
Another bound, known as the \textit{second linear programming bound}, is also given in \cite{mceliece1977new} in the form
\begin{equation}\label{Eq:MRRW_1}
R \lesssim \min_{0 \leq u \leq 1-2\delta} 1+ g(u^2) - g(u^2 + 2\delta u + 2\delta),
\end{equation} where the function $g(x) \deff H_2( (1- \sqrt{1-x})/2 )$. 
For $0.273 \leq \delta \leq 0.5$, the bound \eqref{Eq:MRRW_1} simplifies to that of \eqref{eq:MRRW_2}. For $\delta < 0.273$, the inequality \eqref{Eq:MRRW_1} is strictly tighter than \eqref{eq:MRRW_2}.
\end{remark}

Plugging in $\delta = d/n$ into \eqref{eq:MRRW_2}, we have the following bound:
\begin{equation}\label{eq:MRRW_3}
    M \leq 2^{n H_2\left(1/2 - \sqrt{d/n(1 - d/n)} \right) +o(n)}. 
\end{equation}
Note that, due to the $o(n)$ term, the bound \eqref{eq:MRRW_3} is not tighter than \eqref{eq:Plotkin_sp} when $j= \Theta(\sqrt{n})$. This appears to the contrary of the fact the MRRW bound is tighter than all the other bounds for relative distance $0.273 <\delta <0.5$. However, a tailored treatment of the proof technique may lead to a nontrivial bound as in the derivation of \eqref{eq:EB_sp} from \Tref{Thm:EB}. In Section \ref{subsec:Upper_New_r3}, one such bound is given through an alternative proof of the \Tref{Thm:MRRW} by working with the maximal eigenfunctions of Hamming balls. 

\begin{remark}\label{rem:Levenshtein}
    Another line of work pioneered by Levenshtein \cite{levenshtein1978choosing, levenshtein1992designs, levenshtein1998universal} has also attempted to derive bounds on both $A_2(n,d)$ and $A_q(n,d)$ for general $q$ from Delsarte's LP. 
    In the binary case, for many values of finite $n,d$, Levenshtein’s bound is better than the MRRW bound. In the regime $d = \delta n$ and $n\goto \infty$, the two bounds converge. 
    For general $q$, a recent work \cite{boyvalenkov2018refinements} has proposed refinements of the Levenshtein bound in $q$-ary Hamming spaces and also derived $q$-ary analogs of the MacEliece bound (Equation \eqref{eq:ME_bound}). 
\end{remark}

\subsection{Spectral-Based Upper Bound}
\label{subsec:prelim_Sasha}
One approach to proving the first linear programming bound is the spectral-based technique in \cite{barg2006spectral}, which relies on the analysis of eigenvectors of some finite-dimensional operators related to the Krawtchouk polynomials. 
While the main goal of the work \cite{barg2006spectral} is to establish the MRRW bounds from a spectral perspective, some of the analytical results in it can be used to derive upper bounds on $A(n,d)$ in the large minimum distance regime. 
In particular, while all the other upper bounds on $A(n,d)$ scale superpolynomially in $n$ when $j =\Theta(\sqrt{n})$, a sequence of bounds scaling polynomially in $n$ can be derived from \cite{barg2006spectral}. 
A key result in \cite{barg2006spectral} is the following bound on the size of a binary code with minimum distance $d$. 
\begin{theorem}[\cite{barg2006spectral} Theorem 2, binary case]\label{thm:Barg_binaryBound}
    Let $C$ be an $(n, M, d)$ binary code.  Then
    \begin{equation}\label{eq:bargThm3.2}
M \leq \frac{4(n-k)}{n- \lambda_k}\binom{n}{k}
\end{equation}
for all $k$ such that $\lambda_{k-1} \geq n-2d$, where $\lambda_k$ is the maximal eigenvalue of the $(k+1)\times (k+1)$ self-adjoint matrix $S = (s_{i,j})_{i,j =1}^{k+1}$ defined by  $s_{i,i+1} = s_{i+1,i} = \sqrt{i(n+1-i)}$ for $i=1, 2, \ldots, k$ and $s_{i,j} = 0$ otherwise.
\end{theorem}
Upper and lower bounds on  $\lambda_k$ are also provided. 
\begin{lemma}[\cite{barg2006spectral} Lemma 2, binary case]\label{lem:Barg}
Let $k < n/2$. For all $s = 2, \ldots, k+1$, 
\[
2\sqrt{k(n-k+1)} \geq \lambda_k \geq \frac{2(s-1)}{s}\sqrt{(k-s+2)(n-k+s-1)}.
\]
\end{lemma}

To establish bounds on $A(n,d)$ for the regime $d = n/2 - \Theta(\sqrt{n})$, consider a finite $k\in \N$ and $s \in \mathset{2, \ldots, k+1}$. Letting $n \goto \infty$, we have
\[
    \lambda_k \geq \frac{2(s-1)}{s}\sqrt{(k-s+2)(n-k+s-1)} = \frac{2(s-1)}{s}\sqrt{k-s+2}(1+ o(1))\sqrt{n}.
\]
Thus $\lambda_k \gtrsim \underline{\lambda}_k \sqrt{n},$ where $\underline{\lambda}_k$ is given by
\begin{equation}\label{eq:def_lambda_normailized}
    \underline{\lambda}_k = \max_{2\leq s \leq k+1} \bracenv{ \frac{2(s-1)}{s}\sqrt{k-s+2}}.
\end{equation} Since $\lambda_k$ scales as $\Theta(\sqrt{n})$ for all finite $k$, the bound on  $A(n,d)$ is asymptotically equivalent to 
\begin{equation}\label{eq:And_Sasha_bound}
    A(n,d) = O(n^k) \mbox{ as long as } d \geq \frac{n}{2} - \frac{\underline{\lambda}_{k-1}
}{2}\sqrt{n}.
\end{equation}

By solving \eqref{eq:def_lambda_normailized} for $k =1, 2, 3$, we obtain $\underline{\lambda}_1 = 1,\, \underline{\lambda}_2 = \sqrt{2}, \, \underline{\lambda}_3 = \frac{4}{3}\sqrt{2}$, which, via \eqref{eq:And_Sasha_bound}, lead to 
$A(n,d_1) = O(n^2),\, A(n,d_2) = O(n^3),\, A(n,d_3) = O(n^4)$ as long as $d_1 \geq \frac{n}{2} - \frac{1
}{2}\sqrt{n}, \, d_2 \geq \frac{n}{2} - \frac{\sqrt{2}
}{2}\sqrt{n}, \, d_3 \geq \frac{n}{2} - \frac{2\sqrt{2}}{3}\sqrt{n},$ respectively.
Many more bounds for the large minimum distance regime can be obtained by choosing other $k\in \N$. 
These bounds and the new upper bounds shown in Section~\ref{subsec:Upper_general_r} are both polynomial in $n$ and are tighter than all other known bounds, as discussed in Section~\ref{subsec:PrelimMRRW}. 
In Section~\ref{subsec:Upper_general_r_numerical}, the asymptotic behavior of the two types of bounds when $d = n/2 -\rho \sqrt{n}$ are plotted for $\rho \in (0.5, 9.5)$. 

\section{Main Results - Lower Bounds}\label{sec:Main_lower}
Two polynomial-based cyclic code constructions are given in this section. The first construction, described in Section~\ref{subsec:Main_CyclicConstruciton}, leads to a family of codes where the term $j$ ranges from $\Theta(\sqrt{n})$ to $\Theta(n)$. The second construction, described in Section~\ref{subsec: cyclic_general}, applies to a smaller range of $j$, between $\Theta({n}^{2/3})$ and $\Theta(n)$, but are tighter than the first construction over this range. 
Note that the results in this section are not asymptotic and hold for finite values of $n$, i.e., the first construction only requires $n \geq 15$ and the second one requires $n \geq 63$. 

\subsection{Cyclic Code with High Minimum Distance}\label{subsec:Main_CyclicConstruciton}
We construct a binary cyclic code $C$ with high minimum distance as follows. 
\begin{theorem}
\label{thm_construction}
Let $n= 2^m -1 $, and $m\in \N$ be an even integer with $m \geq 4$. Let $c$ be an integer with $0 \leq c \leq m/2-1$. 
There exists a binary cyclic code $C$ of length $n$, dimension $(c+1/2)m$, and minimum distance 
\begin{equation}\label{eq:cyclic_dmin}
    d \geq 2^{m-1} - 2^{m/2 +c-1} \geq {n}/{2}- 2^{c-1}\sqrt{n}. 
\end{equation}
\end{theorem}
\proof
Consider the finite field $F = {\Bbb F}_{2^m}$ and the subfield $K = \Ftwo < F$. Let $\alpha$ be a primitive root of unity in $F$, and set $\alpha_i = \alpha^{1+ 2^{m/2 +i}}$ for $i =0, 1, 2, \ldots, c$. Consider the binary cyclic code with the generator polynomial
\begin{equation*}
    g(x) = \frac{x^n -1}{\prod_{i=0}^{c}M_{\alpha_i}(x) },
\end{equation*}
where $M_{\beta}(\cdot)$ is the minimal polynomial of $\beta$ over $K$.
Note that the $\alpha_i$'s belong to different conjugacy classes, i.e, 
\begin{align*}
A_i &\deff \mathset{\alpha_i^{2^j} \mid j = 0, 1, 2, \ldots,m-1 } = \mathset{\alpha^{ 2^j + 2^{m/2 + i +j} } \mid j=  0, 1, 2, \ldots, m-1}    
\end{align*}
are disjoint subsets of $F\setminus \mathset{0}$, and $\abs{A_0} = m/2$, $\abs{A_i} = m$ for $i\neq 0$. 
This is ensured by the particular choice of $\alpha_i$'s. 
More specifically, let $P_i = \mathset{ 2^j + 2^{m/2 + i +j} \mbox{ mod } 2^m -1  \mid j=  0, 1, \ldots, m-1}$ be the set of the exponents of $\alpha$ for elements in $A_i$. Each $P_i$ is a cyclotomic coset mod $2$ in $F$ and the length-$m$ binary representation for each $p, p'$ in $P_i$ are cyclic shifts of each other. Let $p_i = 1+ 2^{m/2 +i}$ be the coset representative of $P_i$.
The claim on the size of $\abs{A_i}$ holds by noting that $\abs{A_i} =\abs{P_i} = m$ for $i\neq 0$, and $\abs{A_0} =\abs{P_0} = m/2$.
To claim that $A_i$'s are disjoint, it suffices to show that the cyclotomic cosets $P_i$'s are disjoint. First note that for two cyclotomic cosets $P_i$ and $P_k$, they are either disjoint or identical. Assume for some $i\neq k$, cosets $P_i$ and $P_k$ are identical. Then  $p_i = 1 + 2^{m/2 + i }$ is an element in $P_k$, that is, there is a $p' =  2^\ell + 2^{m/2 + k+\ell} \in P_k$ for which $p_i = p'$ modulo $2^m -1$.  As both $p_i$ and $p'$ are sums of two powers of $2$, we note that neither $m \mid \ell$ and $m \mid (\ell+k-i)$, nor $m \mid (\ell -m/2-i)$ and $m\mid (m/2+k +\ell)$, can happen. Hence $p_i\notin P_k$, and thus $P_i$ and $P_k$ are disjoint.
Thus the degree of the polynomial $g(x)$ is $n -(c+1/2)m$. Hence, the dimension of the code is at least $(c + 1/2)m$.

For the minimum distance, let $t = 2^{m-1} + 2^{m/2 +c-1}+1$. We show next that for $j=t, t+1, \ldots, 2^m -1$, $\alpha^j $ is a root for the generator polynomial $g(x)$. In other words, $P_i \cap \{t, t+1, \dots, 2^m -1\} = \emptyset$, for $i=0, 1, 2, \dots,c$. This is by noting that  the elements in $P_i$, after taking modulo $2^m -1$, can be written as a sum of two powers of two, i.e., $2^\ell+2^j$, where the difference between $\ell$ and $j$ is at least $m/2-c$, and that such a number does not belong to $\{t, t+1, \dots, 2^m -1\}$. Hence, the minimum distance of the code $d$ is at least $2^{m} - t + 1= 2^{m-1} - 2^{m/2 +c-1}$ by BCH bound \cite{hocquenghem1959codes, bose1960class} (see also \cite{macwilliams1977theory, reed2012error}). 
\endproof

Note that the parameters of the codes constructed in \Tref{thm_construction} and the  Delsarte--Goethals codes are both  sitting between those of the first order and the second order Reed--Muller (RM) codes of length $n=2^m$. 
More specifically, RM$(m,1)$ has minimum distance equal to $n/2$ and dimension equal to $m+1$, while RM$(m,2)$ has minimum distance $n/4$, and dimension $1+m+\binom{m}{2}$. 
A comparison between the DG codes and our new construction in the regime $j = \Theta(\sqrt{n})$ can be made as follows. 
Let $c \geq 0$ be a finite integer.  The DG$(m,c)$ code has length $n = 2^m$, minimum distance $d = {n}/{2} - 2^{c-1}\sqrt{n}$ and size $2^{-c} n^{c+2}$. On the other hand, the code parameters given in \Tref{thm_construction} are length $n = 2^m -1$, minimum distance $d \geq {n}/{2} - 2^{c-1}\sqrt{n}$ , and size $(n+1)^{c+1/2}$. 
Therefore the former leads to a stronger lower bound on the size, by  a multiplicative factor of $2^{-c}n^{3/2}$, in the asymptotics.

\subsection{Cyclic Constructions for $j = \Omega(n^{2/3})$}\label{subsec: cyclic_general}
We adopt an approach similar to that in the proof of \Tref{thm_construction} to construct a sequence of cyclic binary codes with $j =n -2d$ scaling as $j = \Omega(n^{2/3})$ in this section. This construction is preferred over that in Section~\ref{subsec:Main_CyclicConstruciton} for all $j = \Omega(n^{2/3})$, and yields a tighter bound on $A(n,d)$ than the DG codes does. 

\begin{theorem}
\label{thm_construction_w3}

Let $n= 2^m -1 $, and $m\in \N$ be a multiple of $6$. Let $c$ be an integer with $1 \leq c \leq m/3-1$.  
There exists a binary cyclic code $C$ of length $n$, dimension 
$k = m\left(\frac{m}{6} +\frac{3c^2}{2} +\frac{5c}{2} +\frac56\right),$
 and minimum distance 
\begin{equation}\label{eq:cyclic_w3_dmin}
    d \geq 2^{m-1} - 2^{\frac{2m}{3} +c-1} -2^{\frac{m}{3}+2c-1} \geq {n}/{2}- 2^{c-1}n^{\frac23} - 2^{2c-1}n^{\frac13}. 
\end{equation}
\end{theorem}
\proof
Consider the finite field $F = {\Bbb F}_{2^m}$ and the subfield $K = \Ftwo < F$. Let $\alpha$ be a primitive root of unity in $F$.
Let $\ell= m/3 -c$, and define three sets consisting of triples of integers, 
\begin{align*}
S_1 &=\mathset{(m/3,m/3,m/3)}, \\
S_2 &= \mathset{(d_1,d_1, d_2)\mid d_1\geq \ell,\, d_2\geq \ell,\; d_1\neq d_2,\; 2d_1 + d_2 = m},\\
S_3 &= \{(d_1, d_2, d_3) \mid d_1 > d_2> d_3\geq \ell \mbox{ or }  d_1 > d_3> d_2\geq \ell,\; d_1 + d_2 + d_3 = m\}.
\end{align*}
A combinatorial argument shows that the sizes of the three sets are  
\begin{align*}
\abs{S_1} = 1, \hspace{3mm}
\abs{S_2} = \floor{\frac{m-3l}{2}}, \hspace{3mm}
\abs{S_3} = \frac{1}{3}\sparenv{
    \binom{m-3l+2}{2} - 3 \floor{\frac{m-3l}{2}} -1
}.
\end{align*}

For $(d,e,f) \in S_1\cup S_2 \cup S_3\deff S$, define $\alpha_{d,e,f} = \alpha^{p_{d,e,f}}\in F$ where $p_{d,e,f} = 2^{m-1} + 2^{e+f-1} + 2^{f-1}$. 
We define a binary cyclic code with the generator polynomial
\begin{equation*}
    g(x) = \frac{x^n -1}{\prod_{(d,e,f) \in S}M_{\alpha_{d,e,f}}(x) \cdot \prod_{i=0}^{m/6 +c}M_{\alpha_i}(x) }.
\end{equation*}
For $i = 0, 1, 2, \dots, c$, define the sets $P_i = \mathset{ 2^j + 2^{m/2 + i +j} \mbox{ mod } 2^m -1  \mid j=  0, 1, 2, \ldots, m-1}$ and $A_i = \mathset{\alpha^p \mid p\in P_i}$ (as in the proof of \Tref{thm_construction}).
Consider sets $A_{d,e,f}, P_{d,e,f}$ as follows: 
\begin{align*}
    A_{d,e,f} &\deff \mathset{ \alpha_{d,e,f}^{2^j} \mid j = 0, 1, 2, \ldots} \\
    P_{d,e,f} &= \{ 2^{m-1+j}+ 2^{e+f-1+j} + 2^{f-1+j}\mbox{ mod } 2^m -1 \mid j=  0, 1, 2, \ldots\}. 
\end{align*}
The set $P_{d,e,f}$ consists of the exponents of $\alpha$ for elements in $A_{d,e,f}$.
Note that  $\abs{P_{d,e,f}} = m/3$ for $(d,e,f) \in S_1$, $\abs{P_{d,e,f}} = m$ for $(d,e,f) \in S_2$, and $\abs{P_{d,e,f}} = m$ for $(d,e,f) \in S_3$. 
Two cyclotomic cosets $P_{d,e,f}$ and $P_{d',e',f'}$ are disjoint as long as $(d,e,f) \neq (d',e',f')$. Since each element of $P_i$ is a sum of two powers of $2$ and each element of $P_{d,e,f}$ a sum of three powers of $2$, we also have $P_i \cap P_{d,e,f}=\emptyset$, for all $i \in \mathset{0, 1, 2, \dots, c}$ and $(d,e,f)\in S$.  
Thus the degree of the polynomial $g(x)$ is 
\begin{align*}
    \deg g(x) &= n - \parenv{m \abs{S_3} + m \abs{S_2} + m/3 \abs{S_1}} - m \parenv{m/6 +c +1/2} \\
    &=  n - m\parenv{
        {m}/{6} +{3c^2}/{2} +{5c}/{2} +{5}/{6}
    }.
\end{align*} Hence, the dimension of the code is at least $m\parenv{
        {m}/{6} +{3c^2}/{2} +{5c}/{2} +{5}/{6}
    }$.

For the minimum distance, we proceed similarly to the steps taken in the proof of \Tref{thm_construction}. Let $t = 2^{m-1} + 2^{m-1-\ell} + 2^{m-1-2\ell}+1$. We show next that for $j=t, t+1, \ldots, 2^m -1$, $\alpha^j$  is a root for the generator polynomial $g(x)$. In other words, $P_i \cap \{t, t+1, \dots, 2^m -1\} = \emptyset$, for $i=0, 1, 2, \dots,m/6 +c$, and $P_{d,e,f} \cap \{t, t+1, \dots, 2^m -1\} = \emptyset$ for $(d,e,f) \in S$.
This is by noting that elements in $P_i$ and $P_{d,e,f}$ are sums of two or three powers of $2$, and the powers differ by at least $\ell$. Such a number can not be found in $\{t, t+1, \dots, 2^m -1\}$. Hence, by the BCH bound  \cite{hocquenghem1959codes, bose1960class} (see also \cite{macwilliams1977theory, reed2012error}), the minimum distance $d$ of the code is at least $2^m - t +1 = 2^{m-1} - 2^{m-1-\ell} - 2^{m-1-2\ell} =2^{m-1} - 2^{2m/3 +c-1} -2^{m/3+2c-1}$. \endproof

For the regime $j = \Theta(n^{2/3})$ we compare the performance of the DG codes and the construction in \Tref{thm_construction_w3}. 
Let $c$ be a positive integer. The DG$(m,r)$ codes with  $r = m/6 +c$ has length $n = 2^m$,  minimum distance $d ={n}/{2} - 2^{c-1}n^{2/3}$,  and size $M = n^2 \parenv{\frac{n}{2}}^{\frac{\log n}{6} +c}$.
The cyclic code described in the proof of \Tref{thm_construction_w3} has length $n = 2^m -1$, minimum distance $d' \geq {n}/{2}- 2^{c-1}n^{\frac23} - 2^{2c-1}n^{\frac13}$, and size $M' \geq n^{\frac{\log n}{6} + \frac{3c^2}{2} +\frac{5c}{2} +\frac56}$.
While the terms $j= n- 2d = 2^{c}n^{\frac23}$ and $j' =n -2d' \leq 2^{c}n^{\frac23} - 2^{2c}n^{\frac13}$ are asymptotically equivalent, i.e., $j \gtrsim j'$, the bound of the size based on the new cyclic construction is stronger. More explicitly, we have $M' \geq 2^{c} n^{\frac{3c^2}{2} + \frac{3c}{2}  -1} \cdot M$, where the degree $\frac{3c^2}{2} + \frac{3c}{2}  -1$ is positive for all $c \in \N$.

\begin{remark}\label{rem:gen_upper_1}
    Codes with minimum distance scaling as ${n}/{2} - \Theta(n^{\frac{2}{3}})$ can be constructed in the manner described in \Tref{thm_construction} too, by choosing $c \approx m/6$. Specifically, if one chooses $c = m/6 + r$ in \Tref{thm_construction}, the code would have dimension $k = (m/6 + r+1/2)m$ and minimum distance $d \geq {n}/{2}- 2^{r-1}n^{\frac{2}{3}}.$ For the same $r$, if one chooses $c= r-1$ in \Tref{thm_construction_w3}, the code has dimension $k' =\left(\frac{m}{6} +\frac{3(r-1)^2}{2} +\frac{5(r-1)}{2} +\frac56\right)m$ and minimum distance $d' \geq {n}/{2}- 2^{r-2}n^{\frac23} - 2^{2r-3}n^{\frac13}.$ 
    For all suffciently large $n$ and $r\geq 2$, the latter construction provides a better trade-off since $k' >k$ and ${n}/{2}- 2^{r-2}n^{\frac23} - 2^{2r-3}n^{\frac13} > {n}/{2}- 2^{r-1}n^{\frac{2}{3}}$.

    The advantage of the second construction is even more evident when one considers the following cases. Taking $c = m/6 + s\sqrt{m}$ for $s>0$ in \Tref{thm_construction}, we have a code $C$ with dimension $k =(m/6 + s\sqrt{m}+1/2)m$ and minimum distance $d \geq {n}/{2}- 2^{s\sqrt{m}-1}n^{\frac{2}{3}}.$ Taking $c=  s\sqrt{m}$ in \Tref{thm_construction_w3}, we have a code $C'$ with dimension $k' =\left(\frac{m}{6} +\frac{3s^2 m}{2} +\frac{5s\sqrt{m}}{2} +\frac56\right)m \geq \left(\frac{1+9s^2}{6}m  +\frac56\right)m$, and minimum distance $d' \geq {n}/{2}- 2^{s\sqrt{m}-1}n^{\frac23} - 2^{2s\sqrt{m}-1}n^{\frac13}.$
    For large $n$, the bounds for the minimum distances $d$ and $d'$ are almost the same, and the dimension $k'$ of the code $C'$ is multiple times larger than $k$, as $k' \approx (1+9s^2)k$.
\end{remark}
\begin{remark}\label{rem:gen_upper_2}
    The constructions used in the proofs of \Tref{thm_construction} and \Tref{thm_construction_w3} draw on the fact that nonzero elements in $F$ can be generated by a primitive root of unity $\alpha$. 
    Any nonzero element $\beta\in F$ can thus be expressed as $\alpha^{p}$ for some $p\in \mathset{0, 1, \dots, 2^m-1}$. 
    Using the binary expansion, $p = b_{m-1} 2^{m-1} + b_{m-2}2^{m-2} + \dots + b_1 2^1+b_0$, roots for $M_{\beta}(x)$ are of the form $\alpha^{p \cdot 2^j} = \alpha^{p'}$ where $p' =  b_{m-1} 2^{m-1+j} + b_{m-2+j}2^{m-2+j} + \dots + b_1 2^{1+j}+b_0 2^j \equiv b_{m-1-j} 2^{m-1} + b_{m-2-j}2^{m-2} + \dots +b_1 2^{1+j}+b_0 2^j + b_{m-1} 2^{j-1} + \dots + b_{m-j}$ after taking modulo $2^m-1$.  
    The length-$m$ binary expressions $p =(b_{m-1} b_{m-2} \dots b_1 b_0)_2, p' = (b_{m-1-j} b_{m-2-j} \dots b_{m-j})_2$ are cyclic shifts of each other. 
    The key idea behind the proof techniques of \Tref{thm_construction} and \Tref{thm_construction_w3} is to leverage the BCH bound with a focus on finding length-$m$ binary sequences $(b_{m-1}, b_{m-2}, \dots ,b_1, b_0)$ and two integers $t_1, t_2$ with $t_1< t_2 \leq 2^m-1$, such that cyclic shifts of the binary sequences do not correspond to values in the range $\mathset{t_1, t_1+1 \ldots, t_2}$. 
\end{remark}
\begin{remark}\label{rem:gen_upper_3}
    According to the discussion in Remark~\ref{rem:gen_upper_2}, the BCH-like construction technique for large minimum distance codes described in Sections~\ref{subsec:Main_CyclicConstruciton} and \ref{subsec: cyclic_general} can be tailored towards deriving lower bounds in \textit{narrower} ranges of $d$. 
    In particular, the construction in Section~\ref{subsec:Main_CyclicConstruciton} admits all $m$-bit binary sequences with exactly two $1$'s spacing at least $\ell= m/2-c$ bits apart (distance is evaluated in a wrap-around manner), and that in Section~\ref{subsec: cyclic_general} all $m$-bit binary sequences with two or three $1$'s spacing at least $\ell= m/3-c$ bits apart. 
    By extending such arguments to consider all $m$-bit binary sequences with Hamming weight between $2$ and $w\in \N$, such that the $1$'s are at least $\ell =m/w-c$ bits apart, one can construct codes with minimum distance scaling as $d = n/2 -O(n^{\frac{w-1}{w}})$. 
\end{remark}

\section{Main Results - Upper Bound} \label{sec:Main_Upper}
\subsection{Harmonic Analysis Approach}
\label{subsec:Upper_Navon}

We adopt a covering argument similar to Navon and Samorodnitsky \cite{navon2009linear} and show upper bounds on the size of any code $C$ with length $n$ and minimum distance $d$ scaling as $d \geq n/2 - \Omega(\sqrt{n})$.
The viewpoint presented in \cite{navon2009linear}, providing an alternative proof to the MRRW bound, is different from the original proof found in \cite{mceliece1977new} which relies on analytical properties of the Krawchouk polynomials, and instead employs Fourier analysis on the group $\Ftwo^n$ as their main tool.

In particular, the authors of \cite{navon2009linear} exploit the expediency of working with the maximal eigenfunctions of Hamming balls. One key finding was that, given any real-valued function $f$ on $\bit^n$ with a small support $B\subset \bit^n$, such that the adjacency matrix of the Hamming cube acts on $f$ by multiplying it pointwise by a large factor, the cardinality of error-correcting codes with minimum distance $d$ can be upper bounded by $n\abs{B}$. 
The applicability will depend on the value of the multiplying factor. By proposing functions $f$ supported on Hamming balls $B = B_r(\zero, n)$ of different radii $r$, one may derive a lower bound of the multiplying factor, formally called the \textit{maximal eigenvalue of adjacency matrix of the subgraph incduced by $B$}.   This makes possible a simple proof of the first linear programming bound.

Let us now state the definition of the \textit{maximal eigenvalue} of a graph. Let $G = (V,E)$ be a (finite, undirected, simple) graph. Let $A_G = (A_{ij})$ denote the $\abs{V} \times \abs{V}$ adjacency matrix of $G$, defined by $A_{ij} = 1$ if $(i, j) \in E$ and $A_{ij} = 0$ otherwise for vertices $i, j \in V$. Note that $A_G$ is symmetric, so its eigenvalues are real, and can be ordered as $\lambda_1 \geq \lambda_2 \geq \ldots \geq \lambda_n$.
For any function $f$ on $\Ftwo^n$, the function $Af$ sums at each point of $\bit^n$ the values of $f$ at its neighbours. That is, the value taken by the function $Af$ at a vertex $x\in \Ftwo^n$, denoted by $(Af)(x)$ or  $Af(x)$, is given by $Af(x) = \sum_{y\in \Ftwo^n: w_H(x, y) =1} f(y)$.
When $B$ is a subset of the cube $\Ftwo^n$, set 
\begin{equation}\label{eq:graphEigen_def}
    \lambda_B \deff \max \left\{ 
    \frac{\Span{Af, f}}{\Span{f, f}} \, \Big\vert \, f:\Ftwo^n \rightarrow \R, \mbox{ supp}(f) \subseteq B
    \right\}, 
\end{equation} where $\Span{f,g} \deff \frac{1}{2^n} \sum_{x\in \Ftwo^n} f(x)g(x)$ for real-valued functions $f,g$ on $\Ftwo^n$.
That is, $\lambda_B$ is the maximal eigenvalue of adjacency matrix of the subgraph of $\bit^n$ induced by $B$. 

Two lemmas were shown in \cite{navon2009linear} to show \eqref{eq:MRRW_2}. 
\begin{lemma}[\cite{navon2009linear} Prop 1.1] \label{lem:NavonProp1.1}
 Let $C$ be a code with block length $n$ and minimal distance $d$. Let $B$ be a
subset of $\bit^n$ with $\lambda_B \geq n -2d +1$. Then $\abs{C} = M \leq n \abs{B}$.
\end{lemma}

\begin{lemma}[\cite{navon2009linear} Lemma 1.4]\label{lem:NavonLem1.4}
Let $B = B_r(\zero, n) \subseteq \bit^n$. The maximal eigenvalue associated with $B$ 
is $\lambda_B \geq 2\sqrt{r(n-r)} - o(n)$.
\end{lemma}

To prove \eqref{eq:MRRW_2}, we note that \Lref{lem:NavonLem1.4} implies that a radius $r^* = n/2 - \sqrt{d(n-d)} +o(n)$ exists such that $\lambda_{B_{r^*}} \geq n -2d +1$. \Lref{lem:NavonProp1.1} in turn shows that any code of length $n$ and minimal distance $d$ has at most $n \abs{B_{r^*}} = n\cdot \Vol(r^*, n)$ codewords. The cardinality of a Hamming ball of radius $r$ is $\Vol(r, n) =  2^{H_2(r/n) n + o(n)}$. Equation \eqref{eq:MRRW_3} follows the above argument, hence yielding equation \eqref{eq:MRRW_2}.  

We note that the above argument can not be used directly to show an upper bound when $d = n/2 - \Theta(\sqrt{n})$. In particular, the $o(n)$ term in \Lref{lem:NavonLem1.4} renders the search for a meaningful $r^*$ impossible, as we would ideally require a subset $B$ with $\lambda_B$ close to $n-2d +1 = \Theta(\sqrt{n})$.

\subsection{Improved Bounds for $d \geq n/2 -\sqrt{n}$}\label{subsec:Upper_New_r3}
We show in this section an approach to lower bound $\lambda_B$ for the Hamming ball $B= B_3(\zero, n)$, which, when coupled with a new proposition stronger than \Lref{lem:NavonProp1.1}, leads to an upper bound scaling as $A(n,d) = O(n^{3.5})$ for $d \geq n/2 -\sqrt{n}$.

First we provide a proposition in place of \Lref{lem:NavonLem1.4} that does not require an $o(n)$ term.
\begin{proposition}\label{prop:eigenR=3}
Let  $B= B_3(\zero, n) \subseteq \bit^n$ be the Hamming ball of radius $3$. The maximal eigenvalue associated with $B$ is $\lambda_B \geq \parenv{ \sqrt{3+ \sqrt{6}} +o(1)}\sqrt{n} \gtrsim 2.334 \sqrt{n}$.
\end{proposition}
\proof
Recall the definition of the maximal eigenvalue in \eqref{eq:graphEigen_def}. We prove the proposition by constructing a function $f$ with support in $B$, and for which ${\Span{Af, f}}/{\Span{f, f}} \approx \sqrt{3+ \sqrt{6}}\sqrt{n}$.
The function $f$ will be symmetric, namely its value at a point will depend only on the Hamming weight of the point. 
With a slight abuse of notation, such a function is fully defined by its values $f(0), f(1), \ldots, f(n)$ at Hamming weights $0, 1, \ldots, n$.

Set $f(0) =1$, $f(j) =0$ for $j \geq 4$, and let 
\begin{equation}\label{eq:f_recurrence}
    \lambda f(i) = Af(i) = i f(i-1) + (n-i) f(i+1)
\end{equation}for $i= 0, 1,2$ (assuming $f(-1) =0$), where $\lambda = t \sqrt{n}$. We have 
\begin{align*}
    f(1) &= \frac{\lambda f(0)}{n} = \frac{t}{\sqrt{n}}, \hspace{5mm}
    f(2) = \frac{\lambda f(1) - 1 f(0)}{n-1} =\frac{t^2 -1}{n-1},\\
    f(3) &= \frac{\lambda f(2) - 2 f(1)}{n-2} =\frac{1}{n-2} \left(\frac{t^2-1}{n-1}t\sqrt{n} - 2\frac{t}{\sqrt{n}}\right).
\end{align*}

We may use the values $f(i)$ and calculate 
\begin{align*}
    2^n \Span{Af, f} &= 2t\sqrt{n}+ t(t^2 - 1)^2 \frac{n\sqrt{n}}{n-1}    = \parenv{2t+ t(t^2 - 1)^2 + o(1)} \sqrt{n} , \\
    2^n \Span{f, f} &= 1 + t^2 + \frac12 \frac{n}{n-1} (t^2-1)^2+  \frac{1}{6} \frac{n-1}{n-2} t^2 \sparenv{
        \frac{n}{n-1} (t^2- 1) -2}^2\\
    &=1 + t^2 + (t^2-1)^2/2 +  t^2 (t^2-3)^2 /6 +o(1).
\end{align*} 
We are now ready to optimize the value 
\begin{equation}\label{eq:EigenPf_1}
    \frac{\Span{Af, f}}{\Span{f, f}}  = \left[ \frac{2t + t(t^2-1)^2 }{(t^6 - 3t^4 +  9t^2 +9)/6 }  + o(1) \right] \sqrt{n}
\end{equation} over $t> 0$.
Taking $t=  \sqrt{3+ \sqrt{6}}$, the square bracket term in \eqref{eq:EigenPf_1} achieves its maximum $ \sqrt{3+ \sqrt{6}} + o(1)$. 
\endproof

In order to provide a bound as tight as possible, we improve upon \Lref{lem:NavonProp1.1} and show the following proposition.
\begin{proposition}\label{prop:tighter_Cbound}
 Let $C$ be a code with block length $n$ and minimal distance $d$. Let $B$ be a
subset of $\bit^n$ with $\lambda_B > n -2d$. Then $\abs{C} = M \leq \frac{n}{\lambda_B - (n-2d)} \abs{B}$.
\end{proposition}
The proof can be shown using a similar argument as in the proof of \Lref{lem:NavonProp1.1} in \cite{navon2009linear}, and is provided in Appendix-\ref{subsec:prop_tighter_Cbound_pf} for reference. 

With Propositions~\ref{prop:eigenR=3} and \ref{prop:tighter_Cbound}, we are ready to state the upper bound on $A(n, \ceil{n/2 - \sqrt{n }\,})$.
\begin{theorem}\label{thm:mainUpper_r3}
 If a $(n, M, d)$ binary code $C$ has minimum distance $d \geq n/2 -\sqrt{n}$, then 
 \begin{equation*}
     M \leq \frac{\sqrt{n}}{\sqrt{3+ \sqrt{6}} -2 +o(1)} \Vol(3,n)= O(n^{3.5}).
 \end{equation*}
\end{theorem}
\proof
Let $B= B_3(\zero, n)$. The maximal eigenvalue induced by $B$ is $\lambda_B \geq \left({\sqrt{3+ \sqrt{6}} +o(1)}\right) \sqrt{n}$ according to \Pref{prop:eigenR=3}. Since $n - 2d \leq 2\sqrt{n} \lesssim \lambda_B$, the cardinality of $C$ can be upper bounded using \Pref{prop:tighter_Cbound} as 
\begin{equation*}
    M \leq \frac{n}{\lambda_B - (n-2d)} \abs{B} \leq \frac{\sqrt{n}}{ \sqrt{3+ \sqrt{6}} -2 +o(1)}\Vol(3,n). \QEDinEq
\end{equation*}

\begin{remark}\label{rem:R3boudn_exact}
We note that the argument above can upper bound the size as $M = O(n^{3.5})$ as long as $(n -2d)/\sqrt{n}$ is strictly smaller than  $\sqrt{3+ \sqrt{6}}$. That is, for any $d \gtrsim n/2 - \rho\sqrt{n}$, for some constant $\rho < \sqrt{3+ \sqrt{6}}/2 \approx 1.167$, 
we have $A(n,d) = O(n^{3.5})$.
\end{remark}

\subsection{Improved Bounds for $d\geq n/2 - \Theta(\sqrt{n})$}\label{subsec:Upper_general_r}

In general, it is possible to generalize the approach in Section~\ref{subsec:Upper_New_r3} that lower bounds $\lambda_B$ for $B = B_r(\zero, n)$ from $r= 3$ to any given $r\in \N$.
Specifically, setting $\lambda = t\sqrt{n}$,  we would need to apply the recurrence equation \eqref{eq:f_recurrence} iteratively to find $f(i)$ as a function of both $t$ and $n$, for $i = 1, \ldots, r$, compute inner products $\Span{Af, f}$ and $\Span{ f, f}$, and solve the optimization problem that maximizes the quotient as in \eqref{eq:EigenPf_1}.
The procedure could be almost intractable for large (but finite) $r$. 

A more feasible approach is given in this section to lower bound the maximal eigenvalue associated with $B = B_r(\zero, n)$.
The approach is comprised of four parts. 
The first part constructs a symmetric function $g$ on $\bit^n$ based on a recursive relation involving $\lambda = t\sqrt{n}$, and a scaled version of $g$ denoted by $\tilde{g}$, both of which are defined independent of $r$ and have support on the entire domain $\bit^n$. 
In the second part, we define a function $f$ which is identical to $g$ on $B$ and $0$ elsewhere, evaluate the quotient seen in equation \eqref{eq:graphEigen_def}, i.e., ${\Span{Af, f}}/{\Span{f, f}}$, and show that it can be expressed concisely as the difference of $\lambda$ and another term involving $\tilde{g}(r)$ and $\tilde{g}(r+1)$, where $\tilde{g}(i)$ depends on $i, n$ and $t$. The quotient can thus be lower bounded by $\lambda$ which guarantees that either $\tilde{g}(r)$ or $\tilde{g}(r+1)$ is $0$. (That is, for a given $r$, one may choose $t$ and $n$ appropriately so that  $\tilde{g}(r)= 0$, or  $\tilde{g}(r+1)=0$, and $t\sqrt{n} $ would be a lower bound of $\lambda_B$.)
In the third part, we introduce two other functions $h$ and $\tilde{h}$ which, broadly speaking, act as the respective proxies of $g$ and $\tilde{g}$. In particular,  as $n$ grows large, the maximal root of $\tilde{g}(k)$ (when viewed as a function of $t$) converges to that of $\tilde{h}(k)$, denoted by $t_h(k)$, a value independent of $n$.
The fourth part concludes the argument by showing that the quantity $\lambda_B/ \sqrt{n}$ is lower bounded by the maximal root of $\tilde{h}(r+1)$, when viewed as a function of $t$, for sufficiently large $n$. Finally we leverage Proposition~\ref{prop:tighter_Cbound} to show $A(n,d) = O(n^{r+1})$ as long as $(n-2d)/{\sqrt{n}} < t_h(r+1) -s$ for some $s>0$. 
Throughout this section, we assume $\lambda = t \sqrt{n}$ for some constant $t>0$. 

\subsubsection*{Part 1}

We first consider a symmetric function $g:\Ftwo^n \rightarrow \R$, and with a slight abuse of notation, write $g(\mathbf{x}) = g(w_H(\mathbf{x}))$ for $\mathbf{x}\in \Ftwo^n$. 
Define $g$ by the initial condition $g(0) =1$, and the recurrence relations 
\begin{equation}\label{eq:g_rec_def}
\lambda g(i) = Ag(i) = i g(i-1) + (n-i) g(i+1) \mbox{ for } i= 0, 1, 2, \ldots, n-1, 
\end{equation}
assuming $g(-1) =0$.
For example, we have 
$    g(1) =  \frac{\lambda}{n},\;
    g(2) =\frac{1}{n-1}\left(\frac{\lambda^2}{n} -1 \right),\; 
    g(3) =\frac{1}{n-2} \left(
    \frac{\lambda^3}{n(n-1)} -  \frac{2 \lambda}{n} - \frac{\lambda}{n-1}
    \right).$

Define a real-valued function $\tilde{g}$ by $\tilde{g}(i) = n^{i/2}g(i)$ for $i= 0, 1, \ldots, n$. 
The values $\tilde{g}(i)$ for $i = 0, 1,2, 3$ are 
    $\tilde{g}(0) = 1,\; \tilde{g}(1) = t, \;
    \tilde{g}(2) =\frac{n}{n-1}(t^2 -1), \;
    \tilde{g}(3) = \frac{n}{n-2} \left(
    \frac{t^3 n}{n-1} -  2t - \frac{tn}{n-1}
    \right).$
\begin{remark}\label{rem:g_tilde_g_scaling}
    For each $i = 0, 1, 2,\ldots,$ both $g(i)$ and $\tilde{g}(i)$ are functions of $t$ and $n$. 
    For a fixed $t$, $\tilde{g}(i)$ scales with $n$ as $O(1)$, and that $g(i) = O(n^{-i/2})$.
\end{remark}

\begin{remark}\label{rem:tilde_g_coef}
   For any finite $k\in \N$, 
   it can be shown that $\tilde{g}(k+1) = 0$ only when $\tilde{g}(k)\neq 0$ and $ \tilde{g}(k-1)  \neq 0$, and that $\tilde{g}(k)$ is a degree-$k$ polynomial in $t$ with leading coefficient $1 +O(n^{-1})$. Specifically, for a given $k\in \N$, 
   \begin{equation*}
    \tilde{g}(k) =
    \begin{cases}
        t^k (1+O(n^{-1})) - g_{k, k-2} t^{k-2} (1+O(n^{-1})) + \ldots +(-1)^{k/2} g_{k, 0}  (1+O(n^{-1})) &\mbox{ for even } k,\\
        t^k (1+O(n^{-1})) - g_{k, k-2} t^{k-2} (1+O(n^{-1})) + \ldots +(-1)^{(k-1)/2} g_{k, 1}  (1+O(n^{-1})) &\mbox{ for odd } k, \vspace{-6mm}
    \end{cases}
\end{equation*} where the coefficients $g_{k, k-2\ell}, 1\leq \ell \leq \floor{\frac{k}{2}}$ are positive integers independent of $n$ and $t$. 
\end{remark}

\subsubsection*{Part 2}
Let $B = B_r(\zero, n)$ for some finite $r$. 
Consider a symmetric function $f$ supported on $B$, defined by $f(i) = g(i)$ for all $i = 0, 1, \ldots, r$, and $f(i) = 0$ for all $i > r$.
First, we have
\[
    2^n\Span{f, f} = f(0)^2 + \binom{n}{1} f(1)^2 + \ldots + \binom{n}{r} f(r)^2  = \sum_{i=0}^r \binom{n}{i}f(i)^2 
    = \sum_{i=0}^r \binom{n}{i}g(i)^2.
\]
Using the observation $g(k) =  O(n^{-k/2})$ and that $f(0)^2 =1$, the sum scales as $2^n\Span{f, f} = \Theta(1)$. Note that $Af(i) = Ag(i) = \lambda g(i)$ for $i = 0, 1, \ldots, r-1$ and $Af(r) = r f(r-1) + (n-r) f(r+1) = r g(r-1)$. 
Hence,  
\begin{align}
    2^n\Span{Af, f} &= Af(0) f(0) + \binom{n}{1} Af(1) f(1) + \ldots + \binom{n}{r} Af(r) f(r) \nonumber \\
    &= \lambda g(0)^2 + \binom{n}{1} \lambda g(1)^2 + \ldots + \binom{n}{r-1} \lambda g(r-1)^2  + \binom{n}{r}  r g(r-1) g(r) \nonumber \\
    &= \lambda 2^n \Span{f, f} + \binom{n}{r} \left( r g(r-1) g(r) -  \lambda g(r)^2 \right )\nonumber \\
    &= \lambda 2^n \Span{f, f} - \binom{n}{r}  g(r) (n-r)g(r+1)  \nonumber 
    \\
    &= \lambda 2^n \Span{f, f} -  \parenv{ n^{-r}\binom{n}{r}} \parenv{1-\frac{r}{n}} \parenv{n^{r/2} g(r) } \parenv{n^{\frac{r+1}{2}} g(r+1)} \sqrt{n}  \nonumber\\
    &= \lambda 2^n \Span{f, f} -  \parenv{ n^{-r}\binom{n}{r}} \parenv{1-\frac{r}{n}} \tilde{g}(r) \tilde{g}(r+1) \sqrt{n}, \nonumber 
\end{align} where the fourth equality holds by evaluating the recursion relation \eqref{eq:g_rec_def} with $i=r$.

The ratio between $\Span{Af, f}$ and $\Span{ f, f}$ is thus 
\begin{equation}\label{eq:eigenratio_root}
    \frac{\Span{Af, f}}{\Span{ f, f}}    
    =\lambda - \frac{1}{2^n\Span{f,f}} \parenv{ n^{-r}\binom{n}{r}} \parenv{1-\frac{r}{n}} \tilde{g}(r) \tilde{g}(r+1) \sqrt{n},
\end{equation} which scales as $\Theta(\sqrt{n})$ since $2^n\Span{ f, f} = \Theta(1)$.
Denote by $t_g(i) = t_g(i, n)$ the maximal root of $\tilde{g}(i) = \tilde{g}(i, n, \lambda =t\sqrt{n})$ when viewed as a function of $t$, for $i = 1, 2, \ldots$. For example, $t_g(1) = 0$ since $ \tilde{g}(1) = t $, $t_g(2) = 1$ since $ \tilde{g}(2) =\frac{n}{n-1}(t^2 -1)$, and $t_g(3)$ is the maximal root of the polynomial $t^3n - 3t n + 2 t=0$. 
Then $\lambda_B \geq \max{(t_g(r)\sqrt{n}, t_g(r+1)\sqrt{n})}$ because the second term in \eqref{eq:eigenratio_root} is $0$ when $\lambda$ is either $t_g(r)\sqrt{n}$ or $t_g(r+1)\sqrt{n}$.

The first two steps  successfully simplify the problem for finding a lower bound on $\lambda_B$ to the following. First solve $g(r)$ and $g(r+1)$ by recursively applying \eqref{eq:g_rec_def}, and then find the maximal roots of $\tilde{g}(r) = 2^{r/2}g(r)$ and $\tilde{g}(r+1)=2^{(r+1)/2}g(r+1)$, which are viewed as functions of $t$. The recursive steps, however, still pose a great challenge when the radius $r$ for $B = B_r(\zero, n)$ is large. For example, $g(3) = \frac{1}{n-2}\left[ 
\frac{t\sqrt{n}}{n-1}\parenv{t^2 - 1} - \frac{2t}{\sqrt{n}}\right]$ and 
$g(4) = \frac{1}{n-3}\Big(\frac{t\sqrt{n}}{n-2}\left[ \frac{t\sqrt{n}}{n-1}\parenv{t^2 - 1} - \frac{2t}{\sqrt{n}} \right] - \frac{3}{n-1}\parenv{t^2 - 1}\Big)$. 
Solving roots of $\tilde{g}(r)$ and $\tilde{g}(r+1)$, which in general are complicated functions in both $n$ and $t$, is an even greater challenge.
The following third step shows that $t_g(i)$ converges to the maximal root of a simpler polynomial in $t$ for all finite $i$ when $n\rightarrow \infty$.

\subsubsection*{Part 3}

Define by $h$ a real-valued function $h: \mathset{0 ,1, \ldots, n}\rightarrow \R$ by setting $h(0) =1$ and the recurrence relation (assuming $h(-1) =0$)
\begin{equation}\label{eq:h_rec_def}
\lambda h(i) =  i h(i-1) + n h(i+1) \mbox{ for } i= 0, 1, 2, \ldots, n-1,
\end{equation} which differs from \eqref{eq:g_rec_def} only in the coefficient of $h(i+1)$. It can be shown that $h(k+1) = 0$ only when $h(k)\neq 0$ and $ h(k-1)  \neq 0$, and that $h(k)$ is either $0$ or scales as $\Theta(n^{-k/2})$ for each finite $k$.
The functions $h$ and $g$ coincide asymptotically for all finite $k$. We state precisely a bound on the ratio between the two in the following lemma. 
\begin{lemma}\label{lem:g_approx_f}
    For any given $k\in \N\cup \mathset{0}$, if  $h(i)\neq 0$ for all $i\leq k-1$ , then 
    \begin{equation*}\label{eq:g_approx_f}
     g(k) =\begin{cases}
     h(k)(1+ O(n^{-1})) &\mbox{ if } h(k) \neq 0,\\
    O(n^{-(k+2)/2}) &\mbox{ if } h(k) = 0.
     \end{cases}
     \end{equation*}
\end{lemma}
\proof
First note that $g(0) = h(0) =1$, $g(1) = h(1) = \frac{\lambda}{n}$ and $g(2) = \frac{1}{n-1}\left(\frac{\lambda^2}{n} -1 \right) = \frac{1}{n}(1 + \frac{1}{n-1})\left(\frac{\lambda^2}{n} -1 \right) = (1+ O(n^{-1}))\frac{1}{n}\left(\lambda h(1) - 1h(0)\right) = h(2)\left(1+ O(n^{-1})\right)$.  Also, $h(2) = 0  \mbox{ if and only if } g(2) = 0$. Hence the lemma holds for $k= 0, 1, 2$.
Assume, for some $k\geq 2$, $h(i) \neq 0$ and $g(i) = h(i)(1+ O(n^{-1}))$ for $i= 0, 1, \ldots, k$. Then $g(i)$ and $h(i)$  both scale as $\Theta(n^{-i/2})$ for $i= 0, 1, \ldots, k$. Equation \eqref{eq:h_rec_def} yields 
\[h(k+1) = {n^{-1}}\left(\lambda h(k)- k h(k-1) \right)\]
and \eqref{eq:g_rec_def} yields
\begin{align*}
    g(k+1) &= {(n-k)}^{-1}\left(\lambda g(k)- kg(k-1) \right) \\
    &=  \parenv{1+ O(n^{-1})} n^{-1} 
    \sparenv{
        \lambda h(k)(1+ O(n^{-1}))- k h(k-1)\parenv{1+ O(n^{-1})})
    },
\end{align*}
which implies  $g(k+1) =h(k+1)(1+ O(n^{-1}))$ when $h(k+1)\neq 0$ since both $\lambda h(k)$ and  $k h(k-1)$ scale as $\Theta(n^{-(k-1)/2})$. When $h(k+1) =0$, $g(k+1) = O(n^{-1}n^{-(k-1)/2}n^{-1}) = O(n^{-(k+3)/2})$.
Hence the lemma holds by the principle of mathematical induction.
\endproof

Consider a real-valued function $\tilde{h}:  \mathset{0, 1, \ldots, n} \rightarrow \R$ defined by $\tilde{h}(i) =n^{i/2}h(i)$. The values $\tilde{h}(i)$ for $i = 0, 1,2, 3, 4$ are 
$    \tilde{h}(0) = 1, \tilde{h}(1) = t, 
    \tilde{h}(2) =t^2 -1, 
    \tilde{h}(3) = t^3 -3t, 
    \tilde{h}(4) = t^4 -6t^2 +3.$
\begin{remark}\label{rem:tilde_h_recurrence}
The function $\tilde{h}$ satisfies the recurrence relation
\begin{equation}\label{eq:tilde_h_rec_rel}
    t\tilde{h}(i)  =i\tilde{h}(i-1)+\tilde{h}(i+1),
\end{equation} which follows from the recurrence relation \eqref{eq:h_rec_def} and that $\tilde{h}(i) =n^{i/2}h(i)$. 
\end{remark}

\begin{remark}\label{rem:tilde_h_coef}
For $k$ a finite positive integer, $h(k)$ depends on both $n$ and $t$, whereas $\tilde{h}(k)$ is independent of $n$ and is a degree-$k$ monic polynomial of $t$.
The polynomial $\tilde{h}(k)$ can be expressed as follows, 
\begin{equation*}
\tilde{h}(k) =
\begin{cases}
    t^k - g_{k, k-2} t^{k-2} + \ldots +(-1)^{k/2} g_{k, 0}   &\mbox{ for even } k,\\
    t^k  - g_{k, k-2} t^{k-2}  + \ldots +(-1)^{(k-1)/2} g_{k, 1}  &\mbox{ for odd } k,
\end{cases}
\end{equation*} where the coefficients $g_{k, k-2\ell}, 1\leq \ell \leq \floor{\frac{k}{2}}$ are the same as those in Remark~\ref{rem:tilde_g_coef}. 
\end{remark}

Denote by $t_h(i)$ the maximal root of $\tilde{h}(i)$ when viewed as a function of $t$, for $i \geq 1$. For example, $t_h(1) = 0$, $t_h(2) = 1$, and $t_h(3) = \sqrt{3}$. 
We now show that, for all finite $k$,  the maximal roots $t_h(i)$ and  $t_g(i)$ are equal when $n\rightarrow\infty$.

\begin{lemma}\label{lem:tg_limit}
    Let $k\in \N$ be finite. Then $\lim_{n\rightarrow\infty} t_g(k,n) = t_h(k)$.
\end{lemma}
\proof
First we prove that for any finite $k\in \N$, $(t_h(1), t_h(2), \ldots, t_h(k))$ is a strictly increasing sequence, using the principle of mathematical induction. For $k= 3$, the sequence is $(0, 1, \sqrt{3})$, which verifies the claim. 
Assume for a finite $k$, the sequence $(t_h(1), t_h(2), \ldots, t_h(k))$ is strictly increasing. Note that $\tilde{h}(k+1) = t\tilde{h}(k) -k\tilde{h}(k-1)$ is negative when $t = t_h(k)$ since $\tilde{h}(k)= 0$ and $\tilde{h}(k-1) >0$. Since the leading term in $\tilde{h}(k+1)$ is $t^{k+1}$ and thus grows to positive infinity for $t$ large enough, we must have $t_h(k+1) > t_h(k)$.

We now show that, the sequence of roots $t_g(k,n)$ of $\tilde{g}(k)$, for $n = 1, 2, \ldots$, converges to $t_h(k)$. 
Let $\delta \in (0, t_h(k) -t_h(k-1))$ be a small constant such that  
$\tilde{h}(k, t) > 0$ for all $t\in (t_h(k), t_h(k) +\delta]$, 
and
$\tilde{h}(k, t) < 0$ for all $ t\in [t_h(k) -\delta,   t_h(k))$. 
Let $\epsilon >0$ be the minimum $\epsilon = \min\left\{
-\tilde{h}(k, t_h(k) - \delta), \tilde{h}(k, t_h(k) +\delta)
\right\}$. 
Leveraging Lemma \ref{lem:g_approx_f}, $\tilde{g}(k, t_h(k) +\delta) \geq \epsilon (1+ O(n^{-1}))$ and $\tilde{g}(k, t_h(k) -\delta) \leq  -\epsilon (1+O(n^{-1}))$. This implies the existence of a root of 
$\tilde{g}(k)$ when viewed as a function of $t$ in the interval $(t_h(k) -\delta, t_h(k) +\delta)$, for all sufficiently large $n$. Note also that, for $t >t_h(k) +\delta$, $\tilde{g}(k, t)$ is strictly positive for all sufficiently large $n$, because $\tilde{h}(k, t) \geq \epsilon$.
Since $\delta >0$ can be chosen arbitrarily small, the root 
$t_g(k,n) $ converges to $ t_h(k)$ as $n$ grows to infinity.
\endproof
\begin{remark}\label{rem:tg_limit_byConti}

It is known that the roots of a polynomial (counting multiplicities and only up to permutation) depend continuously on the coefficients of the polynomial \cite{harris87roots, hirose2020continuity}. Hence, \Tref{lem:g_approx_f} also follows as a direct consequence of Remarks \ref{rem:tilde_g_coef} and \ref{rem:tilde_h_coef}.
\end{remark}

\subsubsection*{Part 4}
We are now ready to state the main result in this section, which admits a practical approach to lower bound $\lambda_B$ for fixed $r$ and sufficiently large $n$. 
\begin{theorem}\label{thm:lambdaB_r_finite}
Let $B = B_r(\zero, n)$  and  $\lambda_B$ be the maximal eigenvalue of adjacency matrix of the subgraph of $\bit^n$ induced by $B$. Then $\lambda_B$ is lower bounded by $(t_h(r+1) +o(1)) \sqrt{n}$.
\end{theorem}
\proof
Equation~\eqref{eq:eigenratio_root} guarantees that  $\lambda_B \geq \max{
\mathset{ t_g(r, n)\sqrt{n}, t_g(r+1, n)\sqrt{n} }
}$. Since  $ t_h(k)$ is strictly increasing in $k$, Lemma~\ref{lem:tg_limit} implies that for large $n$, the bound reduces to $\lambda_B \geq  (t_h(r+1) +o(1)) \sqrt{n}$. \endproof

Theorem~\ref{thm:lambdaB_r_finite} generalizes Proposition~\ref{prop:eigenR=3} by establishing  lower bounds on $\lambda_{B_r}$ for fixed $r$ other than the special case $r=3$.    This in turn yields a sequence of bounds on $\abs{C}$ whose applicability depends on the scaling behaviour of the minimum distance in terms of the blocklength $n$. 
\begin{corollary}\label{coro:mainUpper_r_finite}
If a $(n, M, d)$ binary code $C$ has  distance $d > \frac12 \sparenv{n -(t_h(r+1) -s)\sqrt{n}\,}$ for some $r\in \N$ and $s>0$, and $n$ is sufficiently large, then
 \begin{equation*}
     M \leq \frac{\sqrt{n}}{s+o(1)} \Vol(r,n)= O(n^{r+\frac12}).
 \end{equation*}
\end{corollary}
\proof
Let $B = B_r(\zero, n)$. We have $n-2d < (t_h(r+1) -s)\sqrt{n} \lesssim (t_h(r+1) +o(1)) \sqrt{n} \leq \lambda_B$ due to Theorem~\ref{thm:lambdaB_r_finite}.
The size of the code $C$ can thus be bounded using Proposition~\ref{prop:tighter_Cbound} as 
\begin{align*}
M \leq \frac{n}{\lambda_B - (n-2d)} \abs{B} \leq  \frac{\sqrt{n}}{(t_h(r+1) +o(1)) - (t_h(r+1) -s)}  \abs{B} = \frac{\sqrt{n}}{s+o(1)} \Vol(r,n). \QEDinEq  
\end{align*}


\subsection{Improved Bounds for $d\geq n/2 - \Theta(\sqrt{n})$ - Numerical Results}\label{subsec:Upper_general_r_numerical}
Using similar steps as in the proof of \Pref{prop:eigenR=3}, one may show lower bounds of the maximal eigenvalues associated with Hamming balls of different radii, which could be a daunting procedure even for a radius as small as $5$.
Alternatively, results from Section~\ref{subsec:Upper_general_r} suggest a more feasible approach. 
For $k = 1, 2, \ldots$, we first find $\tilde{h}(k)$  by solving the recursive relations \ref{eq:tilde_h_rec_rel} with the initial conditions $\tilde{h}(0) = 1, \tilde{h}(-1) = 0$, and solve the maximal roots $t_h(k)$ of the polynomials $\tilde{h}(k)$. Theorem~\ref{thm:lambdaB_r_finite} then yields a bound $\lambda_{B_k} \gtrsim t_h(k+1) \sqrt{n}$.   
For example, we have 
$ \tilde{h}(1) = t, \,\tilde{h}(2) =t^2 -1,\,\tilde{h}(3) = t^3 -3t,\, \tilde{h}(4) = t^4 -6t^2 +3,\, \tilde{h}(5) = t^5 - 10 t^3 +15t.$
The corresponding maximal roots are $t_h(1) =0,\, t_h(2) =1,\, t_h(3) = \sqrt{3},\, t_h(4) = \sqrt{3+ \sqrt{6}}\approx 2.334,\, t_h(5) = \sqrt{5 +\sqrt{10}} \approx 2.857.$
Applying Theorem~\ref{thm:lambdaB_r_finite} to $r=3$ shows that $\lambda_{B_3} \gtrsim {\sqrt{3+ \sqrt{6}}}\sqrt{n}$, which recovers Proposition~\ref{prop:eigenR=3} in Section~\ref{subsec:Upper_New_r3}. 
With computer program assistance, we are able to solve  maximal roots  $t_h(k)$ for $k$ as large as $101$.

\begin{figure}
    \centering
     \includegraphics[width=0.5\textwidth]{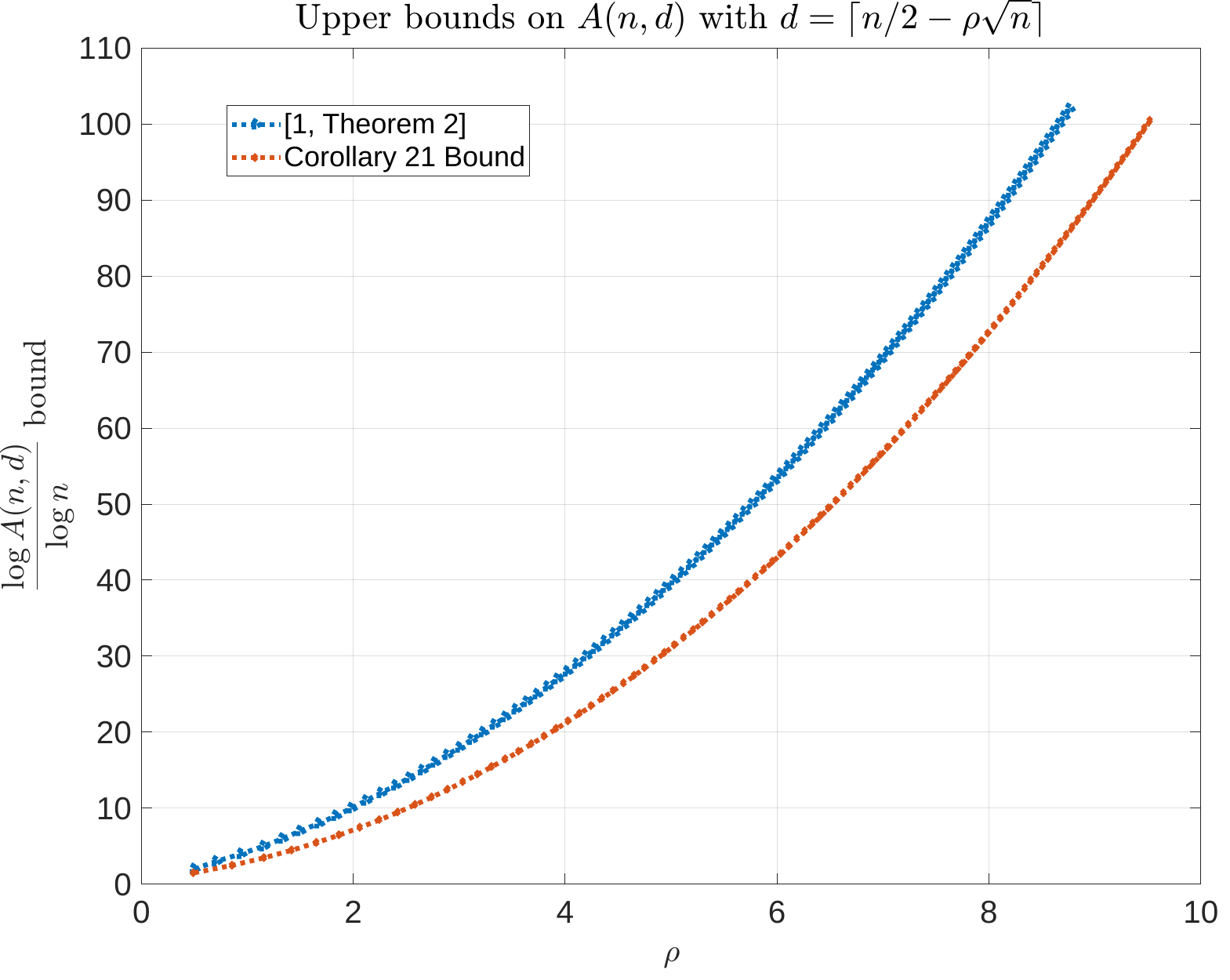}
     \caption{Upper bounds on $\frac{\log A(n,d)}{\log n}$ for $d = \ceil{ n/2 -\rho \sqrt{n}\, }$}
     \label{fig:Mbound_finite_r}
\end{figure}

Consider now a $(n, M, d)$ binary code $C$ with minimum distance $d = \ceil{n/2 -\rho \sqrt{n} \,}$, i.e., $j = 2\rho \sqrt{n}$. 
Corollary~\ref{coro:mainUpper_r_finite} entails the following: If $\rho <{t_h(r+1)}/{2}$ for some $r\in \N$, then $M = O(n^{r+0.5})$. For example, with $\rho=1$, the smallest integer $r$ for the inequality to hold is $r= 3$ since $1 < t_h(4)/2 \approx 1.167$, and thus $M = O(n^{3.5})$. We plot in Figure~\ref{fig:Mbound_finite_r} the the exponent $r+ 0.5$ in the asymptotic bound for $A(n,d)$ for $0.5 =t_h(2)/2 < \rho < t_h(101)/2 \approx 9.5$, based on values of $t_h(k)$ for $k= 2, 3, \dots, 101$.
For example, values of $t_h(3)$ and $t_h(5)$ lead to the bounds 
$A(n, n/2 - \ceil{\rho_1 \sqrt{n}\,}) =O(n^{2.5})$, $A(n, \ceil{n/2 - \rho_2 \sqrt{n}\,}) =O(n^{4.5})$, for all $\rho_1 < \sqrt{3}/2\approx 0.866$ and $\rho_2 < 1.428$. 
One another case, when $r =7$, the point $(2.072, 7.5)$ guarantees that a code with minimum distance $d \geq n/2 -2\,\sqrt{n}$ must have 
$M \lesssim \frac{\sqrt{n}}{4.14 -4} \Vol(7,n) = O(n^{7.5}).$

As discussed in Section~\ref{subsec:prelim_Sasha}, 
the spectral-based bounds for $A(n,d)$ \cite{barg2006spectral} in the large minimum distance regime can be stated as follows: 
\[
A(n,d) = O(n^k) \mbox{ as long as } d \geq {n}/{2} - {\underline{\lambda}_{k-1}
}\sqrt{n}/2,
\]
where $\underline{\lambda}_k$ is defined in equation \eqref{eq:def_lambda_normailized}.
We compute numerically $\underline{\lambda}_k$ for $k= 1,2, \dots, 100$ and show the associated bounds in Figure~\ref{fig:Mbound_finite_r}. We can see that the two families of bounds scale in a similar fashion, while our newly derived upper bounds, due to Corollary~\ref{coro:mainUpper_r_finite}, are slightly tighter. 
\begin{remark}
When the function $f$ is constrained to be symmetric and the support of $f$ constrained to be the Hamming ball $B_r(\zero,n)$, the value $\lambda_B$ (see \eqref{eq:graphEigen_def}) reduces to the maximal eigenvalues $\lambda_r$ in \cite{barg2006spectral}.  
However, due to the distinct proof techniques, our upper bounds, based on harmonic analysis on the Hamming space, only require knowledge of $\lambda_r$, while bounds in \cite{barg2006spectral} require that of both $\lambda_r$ and $\lambda_{r-1}$.
\end{remark}

\section{Conclusion}\label{sec:Conclusion}
In this paper, we study bounds on the cardinality of codes with specified minimum distance $d$ targeting the regime with $d = n/2 - \Omega(\sqrt{n})$. The codes in this regime have vanishing rate, which renders known bounds that dictate the tradeoff between the code rate and relative distance ineffective. We obtain two families of codes based on specifically crafted BCH-like constructions, and a sequence of upper bounds for $d \geq n/2 - \rho \sqrt{n}$ for $\rho \in (0.5, 9.5)$.

The proposed cyclic code constructions are targeted at the regimes $d \geq n/2 - \Omega(\sqrt{n})$ and $d \geq n/2 - \Omega({n}^{2/3})$, and have sizes that are polynomial and quasi-polynomial in $n$, respectively. The proof of the upper bound makes extensive use of Fourier analysis on the Hamming cube as a group, and a new bounding technique for the maximal eigenvalue associated with Hamming balls of finite radii.  

An interesting problem for future work is to study the potential of the Fourier-analytical approach to upper bound the sizes of constant weight codes with large minimum distance. This problem has been studied in the regime $d = \delta n$ with $\delta\in (0, 0.5)$ and the best known result is the second linear programming bound. 
Another interesting problem, as pointed out by authors of \cite{barg2006spectral}, is the underlying similarities between the spectral-based and the Fourier-analytical approaches. For example, if one considers only the Hamming balls $B= B_r(\zero, n)$ among all subsets of $\bit^n$, and requires  functions $f$ to be symmetric, the maximal eigenvalue $\lambda_{B_r}$ appears to be equivalent to the value $\lambda_r$ in \cite{barg2006spectral}. This implies that  our bounds on  $\lambda_{B_r}$ improves on the bounds on  $\lambda_r$ for finite $r$, and that for $r$ scaling sub-linearly in $n$, the latter may yield new bounds on $A(n,d)$ in the regime $d = n/2 - \Theta(n^s)$ for $s\in (0.5, 1)$.

\appendix

\subsection{Harmonic Analysis}\label{subsec:Harmonic}
We compile in this section harmonic analysis preliminaries as in \cite{navon2009linear, kahn1989influence}. See \cite{kahn1989influence} for a more detailed treatment.
Here we list several necessary definitions
and simple facts.

Consider the abelian group structure $\Ftwo^n = (\Z/2\Z)^n$ on the hypercube $\bit^n$.
The characters of the abelian group $\Ftwo^n$ are $\mathset{\chi_z}_{z\in \Ftwo^n}$, where $\chi_z: \bit^n \goto \mathset{-1, 1}$ is given by $\chi_z(x) = (-1)^{\langle x, z\rangle}$ and $\langle x, z\rangle = \sum_{i=1}^n x_i z_i$.

Consider the $\R$-vector space $\mathcal{L}(\Ftwo^n) = \mathset{f: \Ftwo^n \goto \R}$ endowed with the inner
product $\langle \cdot, \cdot \rangle$, associated with the uniform distribution on $\bit^n$:
\begin{equation}\label{eq:harmonic_inner}
    \langle f, g \rangle  = \E_{U_n}f g = \frac{1}{2^n} \sum_{x\in \Ftwo^n} f(x) {g(x)}.
\end{equation}

The set of $2^n$ characters $\mathset{\chi_z}_{z\in \Ftwo^n}$ form an orthonormal basis in the space $\mathcal{L}(\Ftwo^n)$, equipped with uniform probability distribution. That is, for each $z, z' \in \bit^n$, 
$\Span{\chi_z, \chi_{z'}} =  \delta_{z, z'},
$ where $\delta$ is the Kronecker delta function. 
The\textit{ Fourier transform} of a function $f \in \mathcal{L}(\Ftwo^n)$  is the function $\mathscr{F}(f) = \widehat{f} \in  \mathcal{L}(\Ftwo^n)$ given by the
coefficients of the unique expansion of $f$ in terms of the characters:
\begin{equation}\label{eq:harmonic_fhatDef}
    f(x) = \sum_z \widehat{f}(z) \chi_z(x) \mbox{ or equivalently,  } 
    \widehat{f}(z) = \Span{f, \chi_z}.
\end{equation}
One may show that  $\mathscr{F}(\mathscr{F}({f})) = 2^n f$, and $\E f = \widehat{f}(0)$.
For $f, g\in \mathcal{L}(\Ftwo^n)$, the Parseval’s identity holds: $\Span{f,g} = \sum_z \widehat{f}(z)\widehat{g}(z) = 2^n \Span{\widehat{f}, \widehat{g}}$.
A special case of the above equality is the following equality: 
$   \E {f}^2 = \sum_z {\widehat{f}(z)}^2.$

The convolution of $f$ and $g$ is defined by $(f * g)(x) = \E_y{f(y)g(x+y)}$ 
$=\frac{1}{2^n} \sum_{y\in \Ftwo^n} f(y) {g(x+y)}$.
The convolution transforms to dot product: $\widehat{f*g} = \widehat{f}\cdot\widehat{g}$. The convolution operator is commutative and associative. 
For arbitrary functions $f, g, h\in \mathcal{L}(\Ftwo^n)$, the following equality holds: 
\begin{equation}\label{eq:harmonic:conv_commu}
\Span{f*g, h} = \Span{f, g* h}.    
\end{equation}
Also, it can be shown that $\E(f*g) = \E f \cdot \E g$ for all functions $f, g \in \mathcal{L}(\Ftwo^n)$.

In this section and in  Appendix-\ref{subsec:prop_tighter_Cbound_pf}, $L\in \mathcal{L}(\Ftwo^n)$ is a function defined by $L(x) = 2^n$ for $x\in \bit^n$ with $w_H(x) =1$, and $L(x) = 0$ otherwise. 
Let $A$ denote the $2^n \times 2^n$ adjacency matrix of $\Ftwo^n$, such that $Af(x) = (Af)(x) =  \sum_{y\in \Ftwo^n: d_H(x, y) =1} f(y)$.
For any $f \in  \mathcal{L}(\Ftwo^n)$ holds $Af = f * L$ because for $x\in \Ftwo^n$, $Af(x) = \sum_{y:d_H(x,y) =1}f(y)$ $= \sum_{y:w_H(y)=1}f(x+y)$ $= \E_y L(y)f(x+y) = (L*f)(x) = (f*L)(x)$.
The Fourier transform of $L$ is the function $ \mathscr{F}(L) = \widehat{L}$ given by 
$\widehat{L}(z) = \Span{L, \chi_z}$ $= \sum_{x: w_H(x) =1}$ $(-1)^{\Span{x,z}}$ $= n -2\cdot w_H(z)$. 

For $C \subset \Ftwo^n$, let $1_C \in \mathcal{L}(\Ftwo^n)$ be the indicator function of $C$. It can be shown that a code $C$ has minimum distance $d$ if and only if $(1_C * 1_C)(x) = 0$ for all $0 < w_H{(x)} <d$.

\subsection{Proof of \Pref{prop:tighter_Cbound}}\label{subsec:prop_tighter_Cbound_pf}
Let $f_B$ be an eigenfunction supported on $B$ corresponding to its maximal eigenvalue $\lambda_B$. That is $\lambda_B = \Span{Af_B, f_B} / \Span{f_B, f_B}$. It is known that the maximum can be attained with an non-negative function $f_B$, and further we have $Af_B \geq \lambda_B f_B$ (see \cite[p.13-15 and appendix C]{friedman2005generalized}) for details). We write $f = f_B$ and $\lambda = \lambda_B$ interchangeably, and denote the Hamming weight of $x\in \Ftwo^n$ by $\abs{x} = w_H(x)$, in this proof. 
As $f$ is supported on $B$, Cauchy-Schwarz inequality yields the following:
\begin{equation}\label{eq:pf_supportB_Cauchy}
    \E^2 f = \Span{f, 1_B}^2 \leq \E f^2 \cdot \E (1_B)^2 =\E f^2 \cdot \abs{B}/2^n.
\end{equation}

Let $\phi \in \mathcal{L}(\Ftwo^n)$ be a function such that $(\widehat{\phi})^2 = \widehat{\phi * \phi } = 1_C * 1_C$. Equivalently, $\phi * \phi = 2^n \widehat{1_C * 1_C} = 2^n \widehat{1_C}^2$.
Therefore we have 
\begin{equation}\label{eq:pf_improve_1}
    \phi * \phi \geq 0 \mbox{ and } \frac{\E (\phi^2)}{\E^2(\phi)} = \frac{(\phi * \phi)(0)}{\widehat{\phi}^2(0)} = \abs{C}. 
\end{equation}
Now let $F = \phi * f$. We estimate the product $\Span{AF, F}$ in two ways. 
First, 
\begin{align*}
\Span{AF, F} &= \Span{(\phi * f)* L, \phi *f} =    \Span{\phi * \phi * f, f * L} \\
&=\Span{\phi * \phi * f, Af} \geq \Span{\phi * \phi * f, \lambda f} \\
&= \lambda \Span{\phi * f,  \phi * f}= \lambda \Span{F, F} = \lambda \E F^2.
\end{align*}

Second, by Parseval’s identity,
\begin{align*}
    \Span{AF, F} &= 2^n \Span{\widehat{AF}, \widehat{F}} = 2^n\Span{\widehat{L}\cdot\widehat{F}, \widehat{F}} = \sum_z \left(n - 2 \abs{z}\right)\widehat{F}^2(z). 
\end{align*}
Since $\widehat{F} = \widehat{\phi} \cdot \widehat{f}$ and   $(\widehat{\phi})^2(z) = (1_C * 1_C)(z)$,
$\widehat{F}(z) = 0$ for all $0 < \abs{z} < d$. We can estimate $\Span{AF, F}$ by
\begin{align*}
    &\sum_z \left(n - 2 \abs{z}\right)\widehat{F}^2(z) 
    = n \widehat{F}^2(0) + \sum_{z: \abs{z}\geq d} \left(n - 2 \abs{z}\right)\widehat{F}^2(z)\\
    &\leq n \widehat{F}^2(0) + (n-2 d )\sum_{z} \widehat{F}^2(z) 
    = n\E^2 F + (n-2d)\E F^2.
\end{align*}

Combining the two estimates, we have the following inequality: $n \E^2 F$ $\geq \left(\lambda -(n-2d) \right)$ $\E F^2$. Since
\begin{equation*}
\E^2 F = \E^2 (\phi * f) = [\widehat{\phi * f}(0)]^2 =[ \widehat{\phi}(0) \widehat{f}(0)]^2 = \E^2 \phi \E^2 f,
\end{equation*}
\vspace{-7mm}
\begin{align*}
    \E F^2 &= \Span{F, F} = \Span{\phi * f, \phi * f} = \Span{\phi *\phi , f * f}\\
    &\geq 1/2^n (\phi *\phi)(0) (f*f)(0) =  1/ 2^n \E \phi^2 \E f^2,
\end{align*} as $\phi * \phi = 2^n \cdot \widehat{1_C}^2 \geq 0$, 
we now have 
\begin{equation}\label{eq:pf_twoEstimates}
    n \E^2 \phi \E^2 f \geq \left(\lambda -(n-2d) \right) \frac{1}{2^n} \E \phi^2 \E f^2.
\end{equation}
Leveraging equations \eqref{eq:pf_supportB_Cauchy}, \eqref{eq:pf_improve_1}, and \eqref{eq:pf_twoEstimates}, 
the size of any code $C$ with minimum distance $d$ is 
\begin{equation*}
    \abs{C} = \frac{\E \phi^2}{\E^2 \phi} 
    \leq \frac{n}{\lambda -(n-2d)  } \cdot 2^n \frac{\E^2 f}{\E f^2}  
    \leq \frac{n}{\lambda -(n-2d) } \abs{B}.\QEDinEq
\end{equation*}
\bibliographystyle{IEEEtran}
\bibliography{IEEEabrv}

\begin{thebibliography}{10}
\providecommand{\url}[1]{#1}
\csname url@samestyle\endcsname
\providecommand{\newblock}{\relax}
\providecommand{\bibinfo}[2]{#2}
\providecommand{\BIBentrySTDinterwordspacing}{\spaceskip=0pt\relax}
\providecommand{\BIBentryALTinterwordstretchfactor}{4}
\providecommand{\BIBentryALTinterwordspacing}{\spaceskip=\fontdimen2\font plus
\BIBentryALTinterwordstretchfactor\fontdimen3\font minus
  \fontdimen4\font\relax}
\providecommand{\BIBforeignlanguage}[2]{{%
\expandafter\ifx\csname l@#1\endcsname\relax
\typeout{** WARNING: IEEEtran.bst: No hyphenation pattern has been}%
\typeout{** loaded for the language `#1'. Using the pattern for}%
\typeout{** the default language instead.}%
\else
\language=\csname l@#1\endcsname
\fi
#2}}
\providecommand{\BIBdecl}{\relax}
\BIBdecl

\bibitem{barg2006spectral}
A.~M. Barg and D.~Y. Nogin, ``Spectral approach to linear programming bounds on
  codes,'' \emph{Problems of Information Transmission}, vol.~42, no.~2, pp.
  77--89, 2006.

\bibitem{macwilliams1977theory}
F.~J. MacWilliams and N.~J.~A. Sloane, \emph{The theory of error correcting
  codes}.\hskip 1em plus 0.5em minus 0.4em\relax Elsevier, 1977, vol.~16.

\bibitem{guruswami2012essential}
\BIBentryALTinterwordspacing
V.~Guruswami, A.~Rudra, and M.~Sudan, ``Essential coding theory,'' 2022.
  [Online]. Available:
  \url{https://cse.buffalo.edu/faculty/atri/courses/coding-theory/book/}
\BIBentrySTDinterwordspacing

\bibitem{van1998introduction}
J.~H. Van~Lint, \emph{Introduction to coding theory}.\hskip 1em plus 0.5em
  minus 0.4em\relax Springer Science \& Business Media, 1998, vol.~86.

\bibitem{gilbert1952comparison}
E.~N. Gilbert, ``A comparison of signalling alphabets,'' \emph{The Bell system
  technical journal}, vol.~31, no.~3, pp. 504--522, 1952.

\bibitem{varshamov1957estimate}
R.~R. Varshamov, ``Estimate of the number of signals in error correcting
  codes,'' \emph{Docklady Akad. Nauk, SSSR}, vol. 117, pp. 739--741, 1957.

\bibitem{elia1983some}
M.~Elia, ``Some results on the existence of binary linear codes (corresp.),''
  \emph{IEEE Transactions on Information Theory}, vol.~29, no.~6, pp. 933--934,
  1983.

\bibitem{barg2000strengthening}
A.~Barg, S.~Guritman, and J.~Simonis, ``Strengthening the {Gilbert--Varshamov}
  bound,'' \emph{Linear Algebra and its Applications}, vol. 307, no. 1-3, pp.
  119--129, 2000.

\bibitem{o2006bounds}
K.~M. O’Brien and P.~Fitzpatrick, ``Bounds on codes derived by counting
  components in {Varshamov} graphs,'' \emph{Designs, Codes and Cryptography},
  vol.~39, no.~3, pp. 387--396, 2006.

\bibitem{jiang2004asymptotic}
T.~Jiang and A.~Vardy, ``Asymptotic improvement of the {Gilbert-Varshamov}
  bound on the size of binary codes,'' \emph{IEEE Transactions on Information
  Theory}, vol.~50, no.~8, pp. 1655--1664, 2004.

\bibitem{spasov2009some}
D.~Spasov and M.~Gushev, ``Some notes on the binary {Gilbert-Varshamov}
  bound,'' in \emph{Sixth International Workshop on Optimal Codes and Related
  Topics, Varna, Bulgaria}, 2009.

\bibitem{ye2021improving}
Z.~Ye, H.~Zhang, R.~Li, J.~Wang, G.~Yan, and Z.~Ma, ``Improving the
  {Gilbert-Varshamov} bound by graph spectral method,'' \emph{arXiv preprint
  arXiv:2104.01403}, 2021.

\bibitem{tsfasman2013algebraic}
M.~Tsfasman and S.~G. Vladut, \emph{Algebraic-geometric codes}.\hskip 1em plus
  0.5em minus 0.4em\relax Springer Science \& Business Media, 2013, vol.~58.

\bibitem{litsyn1998update}
S.~Litsyn, ``An update table of the best binary codes known,'' \emph{Handbook
  of Coding Theory}, 1998.

\bibitem{agrell}
\BIBentryALTinterwordspacing
E.~Agrell, ``Bounds for unrestricted binary codes.'' [Online]. Available:
  \url{https://codes.se/bounds/unr.html}
\BIBentrySTDinterwordspacing

\bibitem{brouwer}
\BIBentryALTinterwordspacing
A.~Brouwer. [Online]. Available:
  \url{https://www.win.tue.nl/\%7Eaeb/codes/binary-1.html}
\BIBentrySTDinterwordspacing

\bibitem{gaborit2008asymptotic}
P.~Gaborit and G.~Zemor, ``Asymptotic improvement of the {Gilbert}--{Varshamov}
  bound for linear codes,'' \emph{IEEE Transactions on Information Theory},
  vol.~54, no.~9, pp. 3865--3872, 2008.

\bibitem{mceliece1977new}
R.~McEliece, E.~Rodemich, H.~Rumsey, and L.~Welch, ``New upper bounds on the
  rate of a code via the {Delsarte-MacWilliams} inequalities,'' \emph{IEEE
  Transactions on Information Theory}, vol.~23, no.~2, pp. 157--166, 1977.

\bibitem{delsarte1973algebraic}
P.~Delsarte, ``An algebraic approach to the association schemes of coding
  theory,'' \emph{Philips Res. Rep. Suppl.}, vol.~10, pp. vi+--97, 1973.

\bibitem{friedman2005generalized}
J.~Friedman and J.-P. Tillich, ``Generalized {Alon--Boppana} theorems and
  error-correcting codes,'' \emph{SIAM Journal on Discrete Mathematics},
  vol.~19, no.~3, pp. 700--718, 2005.

\bibitem{navon2009linear}
M.~Navon and A.~Samorodnitsky, ``Linear programming bounds for codes via a
  covering argument,'' \emph{Discrete \& Computational Geometry}, vol.~41,
  no.~2, p. 199, 2009.

\bibitem{samorodnitsky2021one}
A.~Samorodnitsky, ``One more proof of the first linear programming bound for
  binary codes and two conjectures,'' \emph{arXiv preprint arXiv:2104.14587},
  2021.

\bibitem{barg2008functional}
A.~Barg and D.~Nogin, ``A functional view of upper bounds on codes,'' in
  \emph{Coding and Cryptology}.\hskip 1em plus 0.5em minus 0.4em\relax World
  Scientific, 2008, pp. 15--24.

\bibitem{barg1999numerical}
A.~Barg and D.~B. Jaffe, ``Numerical results on the asymptotic rate of binary
  codes.'' \emph{Codes and Association Schemes}, vol.~56, pp. 25--32, 1999.

\bibitem{coregliano2021complete}
L.~N. Coregliano, F.~G. Jeronimo, and C.~Jones, ``A complete linear programming
  hierarchy for linear codes,'' \emph{arXiv preprint arXiv:2112.09221}, 2021.

\bibitem{loyfer2022linear}
E.~Loyfer and N.~Linial, ``Linear programming hierarchies in coding theory:
  Dual solutions,'' \emph{arXiv preprint arXiv:2211.12977}, 2022.

\bibitem{loyfer2022new}
------, ``New lp-based upper bounds in the rate-vs.-distance problem for binary
  linear codes,'' \emph{IEEE Transactions on Information Theory}, 2023.

\bibitem{ratasuk2016overview}
R.~Ratasuk, N.~Mangalvedhe, Y.~Zhang, M.~Robert, and J.-P. Koskinen, ``Overview
  of narrowband {IoT} in {LTE} {Rel-13},'' in \emph{2016 IEEE Conference on
  Standards for Communications and Networking}.\hskip 1em plus 0.5em minus
  0.4em\relax IEEE, 2016, pp. 1--7.

\bibitem{shirvanimoghaddam2016raptor}
M.~Shirvanimoghaddam and S.~Johnson, ``Raptor codes in the low {SNR} regime,''
  \emph{IEEE Transactions on Communications}, vol.~64, no.~11, pp. 4449--4460,
  2016.

\bibitem{fereydounian2019channel}
M.~Fereydounian, M.~V. Jamali, H.~Hassani, and H.~Mahdavifar, ``Channel coding
  at low capacity,'' in \emph{IEEE Information Theory Workshop}, 2019, pp.
  1--5.

\bibitem{abbasi2021hybrid}
F.~Abbasi, H.~Mahdavifar, and E.~Viterbo, ``Hybrid non-binary repeated polar
  codes for low-{SNR} regime,'' in \emph{IEEE International Symposium on
  Information Theory}, 2021, pp. 1742--1747.

\bibitem{dumer2020codes}
I.~Dumer and N.~Gharavi, ``Codes for high-noise memoryless channels,'' in
  \emph{IEEE International Symposium on Information Theory and Its
  Applications}, 2020, pp. 101--105.

\bibitem{abbasiy2021polar}
F.~Abbasi, H.~Mahdavifar, and E.~Viterbo, ``Polar coded repetition for
  low-capacity channels,'' in \emph{IEEE Information Theory Workshop}, 2021,
  pp. 1--5.

\bibitem{dumer2021combined}
I.~Dumer and N.~Gharavi, ``Combined polar-{LDPC} design for channels with high
  noise,'' in \emph{IEEE Information Theory Workshop}, 2021, pp. 1--6.

\bibitem{abbasi2022polar}
F.~Abbasi, H.~Mahdavifar, and E.~Viterbo, ``Polar coded repetition,''
  \emph{IEEE Transactions on Communications}, vol.~70, no.~10, pp. 6399--6409,
  2022.

\bibitem{jamali2021massive}
M.~V. Jamali and H.~Mahdavifar, ``Massive coded-{NOMA} for low-capacity
  channels: A low-complexity recursive approach,'' \emph{IEEE Transactions on
  Communications}, vol.~69, no.~6, pp. 3664--3681, 2021.

\bibitem{dumer2021codes}
I.~Dumer and N.~Gharavi, ``Codes approaching the shannon limit with polynomial
  complexity per information bit,'' in \emph{IEEE International Symposium on
  Information Theory}, 2021, pp. 238--243.

\bibitem{vaezi2022cellular}
M.~Vaezi, A.~Azari, S.~R. Khosravirad, M.~Shirvanimoghaddam, M.~M. Azari,
  D.~Chasaki, and P.~Popovski, ``Cellular, wide-area, and non-terrestrial
  {IoT}: A survey on {5G} advances and the road toward {6G},'' \emph{IEEE
  Communications Surveys \& Tutorials}, vol.~24, no.~2, pp. 1117--1174, 2022.

\bibitem{abbasi2022hybrid}
F.~Abbasi, H.~Mahdavifar, and E.~Viterbo, ``Hybrid non-binary repeated polar
  codes,'' \emph{IEEE Transactions on Wireless Communications}, vol.~21, no.~9,
  pp. 7582--7594, 2022.

\bibitem{sidel1971mutual}
V.~M. Sidel'nikov, ``On mutual correlation of sequences,'' in \emph{Doklady
  Akademii Nauk}, vol. 196, no.~3.\hskip 1em plus 0.5em minus 0.4em\relax
  Russian Academy of Sciences, 1971, pp. 531--534.

\bibitem{tietavainen1980bounds}
A.~Tiet{\"a}v{\"a}inen, ``Bounds for binary codes just outside the {Plotkin}
  range,'' \emph{Information and Control}, vol.~47, no.~2, pp. 85--93, 1980.

\bibitem{krasikov1997upper}
I.~Krasikov and S.~Litsyn, ``On upper bounds for the distance of codes of small
  size,'' in \emph{Proceedings of IEEE International Symposium on Information
  Theory}, 1997, p.~84.

\bibitem{kasami1968new}
T.~Kasami, S.~Lin, and W.~Peterson, ``New generalizations of the reed-muller
  codes--i: Primitive codes,'' \emph{IEEE Transactions on Information Theory},
  vol.~14, no.~2, pp. 189--199, 1968.

\bibitem{delsarte1970generalized}
P.~Delsarte, J.-M. Goethals, and F.~J. Mac~Williams, ``On generalized
  reedmuller codes and their relatives,'' \emph{Information and control},
  vol.~16, no.~5, pp. 403--442, 1970.

\bibitem{goethals1974two}
J.~Goethals, ``Two dual families of nonlinear binary codes,'' \emph{Electronics
  Letters}, vol.~10, no.~23, pp. 471--472, 1974.

\bibitem{goethals1976nonlinear}
J.-M. Goethals, ``Nonlinear codes defined by quadratic forms over {GF(2)},''
  \emph{Information and control}, vol.~31, no.~1, pp. 43--74, 1976.

\bibitem{hergert1990delsarte}
F.~B. Hergert, ``On the {D}elsarte-{G}oethals codes and their formal duals,''
  \emph{Discrete mathematics}, vol.~83, no. 2-3, pp. 249--263, 1990.

\bibitem{plotkin1960binary}
M.~Plotkin, ``Binary codes with specified minimum distance,'' \emph{IRE
  Transactions on Information Theory}, vol.~6, no.~4, pp. 445--450, 1960.

\bibitem{levenshtein1978choosing}
V.~Levenshtein, ``On choosing polynomials to obtain bounds in packing
  problems,'' in \emph{Proc. Seventh All-Union Conf. on Coding Theory and
  Information Transmission, Part II, Moscow, Vilnius}, 1978, pp. 103--108.

\bibitem{levenshtein1992designs}
V.~I. Levenshtein, ``Designs as maximum codes in polynomial metric spaces,''
  \emph{Acta Applicandae Mathematica}, vol.~29, no.~1, pp. 1--82, 1992.

\bibitem{levenshtein1998universal}
------, ``Universal bounds for codes and designs,'' \emph{Handbook of Coding
  Theory}, vol.~1, pp. 499--648, 1998.

\bibitem{boyvalenkov2018refinements}
P.~Boyvalenkov, D.~Danev, and M.~Stoyanova, ``Refinements of {Levenshtein}
  bounds in q-ary hamming spaces,'' \emph{Problems of Information
  Transmission}, vol.~54, no.~4, pp. 329--342, 2018.

\bibitem{hocquenghem1959codes}
A.~Hocquenghem, ``Codes correcteurs d'erreurs,'' \emph{Chiffers (in French)},
  vol.~2, pp. 147--156, 1959.

\bibitem{bose1960class}
R.~C. Bose and D.~K. Ray-Chaudhuri, ``On a class of error correcting binary
  group codes,'' \emph{Information and control}, vol.~3, no.~1, pp. 68--79,
  1960.

\bibitem{reed2012error}
I.~S. Reed and X.~Chen, \emph{Error-control coding for data networks}.\hskip
  1em plus 0.5em minus 0.4em\relax Springer Science \& Business Media, 2012,
  vol. 508.

\bibitem{harris87roots}
\BIBentryALTinterwordspacing
G.~Harris and C.~Martin, ``Shorter notes: The roots of a polynomial vary
  continuously as a function of the coefficients,'' \emph{Proceedings of the
  American Mathematical Society}, vol. 100, no.~2, pp. 390--392, 1987.
  [Online]. Available: \url{http://www.jstor.org/stable/2045978}
\BIBentrySTDinterwordspacing

\bibitem{hirose2020continuity}
K.~Hirose, ``Continuity of the roots of a polynomial,'' \emph{The American
  Mathematical Monthly}, vol. 127, no.~4, pp. 359--363, 2020.

\bibitem{kahn1989influence}
J.~Kahn, G.~Kalai, and N.~Linial, \emph{The influence of variables on {Boolean}
  functions}.\hskip 1em plus 0.5em minus 0.4em\relax Citeseer, 1989.

\end{thebibliography}


\begin{thebibliography}{1}
\providecommand{\url}[1]{#1}
\csname url@samestyle\endcsname
\providecommand{\newblock}{\relax}
\providecommand{\bibinfo}[2]{#2}
\providecommand{\BIBentrySTDinterwordspacing}{\spaceskip=0pt\relax}
\providecommand{\BIBentryALTinterwordstretchfactor}{4}
\providecommand{\BIBentryALTinterwordspacing}{\spaceskip=\fontdimen2\font plus
\BIBentryALTinterwordstretchfactor\fontdimen3\font minus
  \fontdimen4\font\relax}
\providecommand{\BIBforeignlanguage}[2]{{%
\expandafter\ifx\csname l@#1\endcsname\relax
\typeout{** WARNING: IEEEtran.bst: No hyphenation pattern has been}%
\typeout{** loaded for the language `#1'. Using the pattern for}%
\typeout{** the default language instead.}%
\else
\language=\csname l@#1\endcsname
\fi
#2}}
\providecommand{\BIBdecl}{\relax}
\BIBdecl

\bibitem{ding2017survey}
Z.~Ding, X.~Lei, G.~K. Karagiannidis, R.~Schober, J.~Yuan, and V.~K. Bhargava,
  ``A survey on non-orthogonal multiple access for {5G} networks: Research
  challenges and future trends,'' \emph{IEEE J. Sel. Areas Commun.}, vol.~35,
  no.~10, pp. 2181--2195, 2017.

\end{thebibliography}

\end{document}